\documentclass[twocolumn,trackchanges, tighten, times]{aastex631} 
\usepackage{multirow}
\usepackage{amsmath}
\usepackage{xcolor}
\usepackage{gensymb}
\usepackage{csquotes}
\usepackage{mathtools}
\usepackage{enumitem}
\usepackage{fancybox}
\usepackage{graphicx}
\usepackage{amssymb}
\usepackage{ natbib, bm}
\usepackage{orcidlink}
\usepackage{tabularx}
\usepackage{booktabs}

\graphicspath{{plots/}}

\providecommand{\pgfsyspdfmark}[3]{}
\usepackage[english]{babel}

\newcommand{\mhalo}{{M}_{\rm{halo}}}

\newcommand{\mstar}{{M}_{\star}}

\newcommand{\msun}{{\rm M}_{\odot}}

\newcommand{\lt}{<}
\newcommand{\gt}{>}
\newcommand{\z}{{\it z}}

\newcommand{\jwst}{\textit{JWST}}

\DeclareRobustCommand{\ion}[2]{%
\relax\ifmmode
\ifx\testbx\f@series
{\mathbf{#1\,\mathsc{#2}}}\else
{\mathrm{#1\,\mathsc{#2}}}\fi
\else\textup{#1\,{\mdseries\textsc{#2}}}%
\fi}

\begin{document}

\title[What becomes of massive high-\boldmath{$z$} galaxies?]
{What becomes of \textit{JWST}/NIRCam-selected high-redshift massive galaxies?} 
\correspondingauthor{Devontae Baxter}
\email{dcbaxter@princeton.edu}

\author[0000-0002-8209-2783]{Devontae C. Baxter}
\altaffiliation{NSF Astronomy and Astrophysics Postdoctoral Fellow \\}
\affiliation{Department of Astronomy \& Astrophysics,
University of California, San Diego, 9500 Gilman Dr, La Jolla, CA 92093, USA \\}
\affiliation{Department of Astrophysical Sciences, Princeton University, 4 Ivy Lane, Princeton, NJ 08544, USA\\}

\author[0000-0002-2583-5894]{Alison L. Coil}
\affiliation{Department of Astronomy \& Astrophysics,
University of California, San Diego, 9500 Gilman Dr, La Jolla, CA 92093, USA \\}

\author[0000-0002-1182-3825]{Ethan O. Nadler}
\affiliation{Department of Astronomy \& Astrophysics,
University of California, San Diego, 9500 Gilman Dr, La Jolla, CA 92093, USA \\}

\author[0000-0002-6196-823X]{Xuejian Shen}
\affiliation{Department of Physics \& Kavli Institute for Astrophysics and Space Research, Massachusetts Institute of Technology, Cambridge, MA 02139, USA \\}

\author[0000-0001-8593-7692]{Mark Vogelsberger}
\affiliation{Department of Physics \& Kavli Institute for Astrophysics and Space Research, Massachusetts Institute of Technology, Cambridge, MA 02139, USA \\}

\begin{abstract}
Early \textit{JWST}/NIRCam surveys revealed a puzzling population of high-redshift massive galaxy candidates largely absent from previous rest-frame UV surveys. Spectroscopic follow-up has both confirmed and contested these candidates, whose potential overabundance may challenge traditional models of galaxy formation. In this study we evaluate the reliability of the photometric selections used to identify these candidates in observational data by applying them to galaxies in the TNG300 simulation with synthetic dust-attenuated photometry. Among the five observational selection criteria considered, we find that the selection presented by Pérez-González et al. is the most reliable and inclusive. Nevertheless, only 1 of the 18 galaxies at $z\sim5$ with $\mstar \geq 10^{11}~\msun$ in the simulation satisfies this selection; the remaining 17 galaxies are on average $\sim0.5$ mag bluer than the color selection under the adopted dust model. We construct an improved \textit{JWST}/NIRCam color-magnitude selection that provides a more complete census of massive galaxies at $z\sim5$ in TNG300 while excluding dusty, low-mass galaxies identified by criteria from the observational literature. We investigate the descendants of the most massive NIRCam-selected galaxies at $z=7,~4,$ and $2$ in TNG300, finding that they rarely evolve into the most massive galaxies by $z=0$. In general only the high-redshift massive galaxies that undergo substantial late-time ($z\lesssim0.2$) merger-driven growth become the most massive galaxies in the Universe today. Together these results suggest that current observational \textit{JWST}/NIRCam selections are not identifying the most massive high-redshift galaxies, and caution against interpreting high-redshift massive galaxies as direct progenitors of the most massive galaxies at $z=0$. 
\end{abstract}  


\section{Introduction} 
\label{sec:intro}

The unparalleled near-infrared capabilities of \textit{JWST} have ushered in a revolutionary era in high-redshift astrophysics marked by major advances in galaxy evolution and cosmology. These include improved constraints on cosmological parameters \citep{Pascale25, Tdcosmo25}, the evolution of the galaxy stellar mass function \citep{NavarroCarrera24, Weibel24, Shuntov26}, and the drivers of reionization \citep[e.g.,][]{Endsley23, Simmonds24}, as well as unprecedented probes of the earliest stages of galaxy cluster assembly \citep[e.g.,][]{Morishita23, Saxena25, Laishram26, Witten26a} and galaxy quenching \citep[e.g.,][]{Carnall23b, Looser24, Long24, Nanayakkara24, Setton24, deGraaff25, Weibel25}, and the discovery of the most distant galaxies within the first billion years of cosmic history \citep[e.g.,][]{CurtisLake23, Robertson23, Naidu26}. 

The \textit{JWST} era has also produced puzzling discoveries, including the unexpectedly high abundance of high-redshift massive galaxy candidates identified in early \textit{JWST}/NIRCam surveys that were undetected in previous rest-frame UV surveys \citep{Barrufet23, Carnall23, Labbe23, Rodighiero23, Akins23, Barro24}. These observations have inspired extensive follow-up efforts aimed at understanding the nature and origin of these high-redshift massive galaxy candidates.

On the theoretical front, proposed explanations for the potential overabundance of these NIRCam-selected high-$z$ massive galaxies range from enhanced star formation efficiencies in early galaxies \citep[e.g.,][]{Dekel23, Andalman25, WangZihao25, BoylanKolchin25, Shen26}, top-heavy stellar initial mass functions \citep[IMF; e.g.,][]{Lu25, Ziegler25, vanDokkum24}, modifications to early dark energy \citep{Shen24}, and heavy supermassive black hole (SMBH) seeding scenarios that accelerate early galaxy assembly \citep[e.g.,][]{Mould26, Huang24, Yuan24}. Related studies argue that the stellar masses of these candidates are overestimated due to systematic uncertainties and biases in SED fitting, including the outshining of older stellar populations by recent bursts of star formation \citep[e.g.,][]{Endsley23, Narayanan24, Choe26, Lapasia26}, while others suggest that some candidates may instead be low-redshift interlopers masquerading as massive high-redshift galaxies \citep{Forrest24a}.

Complementing theoretical efforts, observational follow-up has sought to confirm the physical properties of these high-redshift massive galaxy candidates. This has included the incorporation of \textit{JWST}/MIRI photometry to provide improved constraints on their stellar populations and dust emission, which has generally resulted in lower stellar mass estimates for the NIRCam-selected candidates \citep{Chworowsky24, Williams24, WangTao25}. Spectroscopic follow-up has also revealed that a subset of these objects belong to a recently discovered population of extremely red, compact objects with characteristic V-shaped SEDs not observed at low redshift, dubbed \textquote{little red dots} \citep[LRDs;][]{Matthee24}. This misidentification of LRDs as high-redshift massive galaxies arises from contamination by active galactic nucleus (AGN) accretion-disk emission, which can systematically inflate inferred stellar masses \citep{Kocevski23, Killi24, Matthee24, Greene24, Berger25, Akins25}. Nevertheless, spectroscopic observations have confirmed that at least some of the NIRCam-selected massive galaxy candidates are both genuinely massive and reside at very high redshift \citep[e.g.,][]{ArrabalHaro23, Lambrides24, Xiao24, Xiao26b, Lapasia26}.

The confirmation of a subset of these high-redshift massive galaxy candidates naturally raises an important question: what do these galaxies become by the present day? More broadly, these discoveries motivate a deeper exploration of the demographics of galaxies selected by the various \textit{JWST}/NIRCam photometric criteria adopted throughout the observational literature. Addressing both questions requires a theoretical approach that selects simulated high-redshift galaxies using criteria directly comparable to those used in observations and traces their evolution to $z=0$. We adopt this approach in this work by using synthetic dust-attenuated \textit{JWST}/NIRCam photometry for galaxies in the IllustrisTNG simulation and applying the same photometric selections used in the observational literature. The goals of this analysis are to evaluate the reliability of these photometric criteria in selecting simulated high-redshift massive galaxies and to determine how these galaxies evolve to the present day, including whether the most massive NIRCam-selected galaxies ultimately become the most massive galaxies by $z=0$. This study contributes to a growing literature using cosmological simulations and semi-analytic models to make predictions and interpret recent \textit{JWST} observations, including studies of the related but distinct population of unexpectedly luminous galaxies at $z\gtrsim10$ \citep[e.g.,][]{Sun23, McCaffrey23, Chaikin26b, Hardin26, Samuel26, Zhao26, Kim26} and high-redshift massive quenched galaxies \citep[e.g.,][]{Lovell23, Xie24, Lagos24, DeLucia26}.

This paper is structured as follows. In \S\ref{sec:Data}, we describe the simulated galaxy sample and the dust radiative transfer procedure used to generate dust-attenuated \jwst/NIRCam photometry. In \S\ref{sec:NIRCam_selection_results} and \S\ref{sec:evolution_results}, we present our main results. We first evaluate the reliability of five NIRCam photometric selections from the literature in identifying simulated high-redshift massive galaxies (\S\ref{subsec:3.1}), then compare the baryonic and dark matter properties of NIRCam-selected galaxies with the coeval progenitors of galaxies with $\mstar^{z=0}>10^{12}~\msun$ (\S\ref{subsec:3.2}), and finally introduce improved NIRCam selection criteria optimized for high-purity identification of massive $z\sim5$ galaxies (\S\ref{subsec:3.3}). In \S\ref{subsec:4.1}, we compare the stellar and halo masses of NIRCam-selected galaxies and their $z=0$ descendants, followed by a comparison of their large-scale environments in \S\ref{subsec:4.2}. We then investigate the stellar mass growth histories of the most massive high-redshift galaxies (\S\ref{subsec:4.3}), followed by an examination of the factors that determine whether they evolve into the most massive galaxies at $z=0$ (\S\ref{subsec:4.4}). In \S\ref{sec:Discussion}, we discuss the implications of our results for selecting progenitors of the most massive halos and galaxies at $z=0$, including protoclusters and proto-brightest cluster galaxies and explore the impact of dust-modeling assumptions on the interpretation of our results. Lastly, in \S\ref{sec:Conclusion}, we summarize the main takeaways from our analysis.

Throughout this study we adopt a cosmology consistent with \citet{PlanckCollab16}: $\Omega_{\rm{m}}=0.3089$, $\Omega_{\Lambda}=0.6911$, $\Omega_{\rm{b}}=0.0486$, $h=0.6774$, $\sigma_{8}=0.8159$, and $n_{\rm{s}}=0.9667$. We assume a \citet{Chabrier03} initial mass function, and all magnitudes are reported on the AB system \citep{OkeGunn83}. We define stellar masses as the sum of the masses of all stellar particles gravitationally bound to an individual subhalo.

\section{Simulation Data and Synthetic Dust-attenuated NIRCam Photometry}
\label{sec:Data}

\subsection{IllustrisTNG Simulation}
\label{subsec:2.1}

We construct a simulated galaxy sample from \textit{The Next Generation Illustris Project}\footnote{\href{https://www.tng-project.org}{https://www.tng-project.org}} \citep[IllustrisTNG,][]{Nelson18, Naiman18, Springel18, Pillepich18, Marinacci18}, a suite of magnetohydrodynamical cosmological simulations run with the moving-mesh \texttt{AREPO} code \citep{Springel10} within the $\Lambda$CDM paradigm. We use the TNG300-1 run, which is the highest-resolution realization of the large-volume simulations, with $2 \times 2500^{3}$ resolution elements (dark matter + baryons) within a periodic cubic volume of side length $\sim300~{\rm cMpc}$. TNG300-1 has a gravitational softening length of $1.5$ kpc for stars and dark matter at $z=0$ and a minimum softening length of $0.37~\rm{ckpc}$ for gas cells. This results in an average baryonic (dark matter) particle mass resolution of $m_{\rm baryon} = 1.1 \times 10^{7}~\msun$ ($m_{\rm DM} = 5.9 \times 10^{7}~\msun$). 

The simulation uses the fiducial TNG model which includes prescriptions for the formation, growth, and mergers of SMBHs (including dual-mode SMBH feedback), star formation in the dense interstellar medium, stellar population evolution and chemical enrichment, supernova-driven galactic-scale outflows, and gas radiative cooling. Additionally, the physical model was calibrated to reproduce various observations at $z=0$ including the galaxy stellar mass function, stellar-to-halo mass relation, stellar size and BH–galaxy mass relations, as well as the shape of the cosmic star formation rate density at $z \lesssim 10$ \citep{Weinberger17, Pillepich18b}.

Dark matter halos in TNG300 are identified using the friends-of-friends \citep[FoF;][]{Davis85} algorithm with a linking length $b=0.2$. The substructure within the halos (i.e., subhalos) are identified using the \texttt{SUBFIND} algorithm \citep{Springel01, Dolag09}. The properties of (sub)halos are tracked across 100 snapshots spanning from $z=20$ to $z=0$ using merger trees constructed from the \texttt{SUBLINK} algorithm \citep{RodriguezGomez15}. Our analysis makes use of both the main descendant and progenitor branches of the \texttt{SUBLINK} merger trees to investigate the temporal evolution of the (sub)halos identified in this study.   

\begin{table*}
\centering
\footnotesize 
\caption{Summary of \textit{JWST}/NIRCam color-magnitude selection criteria explored in this study.}
\setlength{\tabcolsep}{2.25pt} 
\begin{tabularx}{\textwidth}{cccc}
\hline
Selection & NIRCam Selection Criteria & Redshift Range & Reference \\
\hline

S1 &  $(m_{\rm{150W}} - m_{\rm{356W}}) > 1.5$ $\wedge$ $m_{\rm{356W}} < 27.5$ & $2 \lesssim z \lesssim 6$ & \citealt{PerezGonzalez23} \\

S2 & $m_{\rm{444W}} < 25$ $\wedge$ $m_{\rm{150W}} > 26.5$ & $3 \lesssim z \lesssim 7.5$ & \citealt{GomezGuijarro23} \\

S3 & $(m_{\rm{150W}} - m_{\rm{444W}}) > 2.1$ $\wedge$ $m_{\rm{150W}} > 25$ & $3 \lesssim z \lesssim 8$ & \citealt{Gottumukkala24} \\

S4 & $m_{\rm{444W}} < 27$ $\wedge$ $m_{\rm{150W}} < 29$ $\wedge$ $(m_{\rm{150W}} - m_{\rm{277W}}) < 0.7$ $\wedge$ $(m_{\rm{277W}} - m_{\rm{444W}}) > 1.0$ & $7.4 \lesssim z \lesssim 9.1$ & \citealt{Labbe23} \\

S5 & $(m_{\rm{277W}} - m_{\rm{444W}}) > 1.8$ & $z\gtrsim 6-7$ & \citealt{Akins23} \\

\hline
\end{tabularx}
\label{tab:NIRCam_selections}
\end{table*}

\subsection{Synthetic Dust-Attenuated NIRCam Photometry}
\label{subsec:2.2}

In order to replicate \textit{JWST}/NIRCam selections used in the literature on the TNG300 simulation, we incorporate the synthetic dust-attenuated \textit{JWST}/NIRCam photometry presented in \citet{Vogelsberger20} and \citet{Shen20} for galaxies in the IllustrisTNG simulation from $z=10$ to $z=2$. These data enable us to apply photometric selection criteria from the observational literature that have been used to identify massive, high-redshift galaxy candidates and to assess the effectiveness of these selections in recovering high-redshift massive galaxies in TNG300. Specifically, we use the dust-attenuated apparent band magnitudes derived from performing full dust Monte Carlo radiative transfer calculations of photons propagating through the dusty interstellar medium of simulated galaxies (referred to as Model C in \citealt{Vogelsberger20}). We provide a comprehensive overview of the modeling procedure used to estimate these magnitudes and refer the reader to Section 3 of \citet{Vogelsberger20} for a more detailed description.

The dust-attenuated magnitudes used here are estimated by combining the distribution of stellar particles and gas in simulated galaxies with a modified version of the publicly available \texttt{SKIRT} code \citep{Baes11, Camps13, CampsBaes15, Saftly14} to perform full Monte Carlo continuum dust radiative transfer calculations. This modified version of \texttt{SKIRT} replaces the default SED templates with a family of templates obtained from calculating dust-free, rest-frame SEDs from a two-dimensional grid comprised of initial stellar metallicities and stellar ages. For each grid point a Simple Stellar Population is constructed along with the corresponding rest-frame SED using the Flexible Stellar Population Synthesis (\texttt{FSPS}) code \citep{Conroy09, ConroyGunn10} with MIST isochrones \citep{Paxton11, Paxton13, Paxton15, Choi16, Dotter16} and the MILES stellar library \citep{Sanchez-Blazquez06}, assuming a Chabrier initial mass function \citep{Chabrier03}. Using this framework \texttt{SKIRT} generates an SED for all star particles based on their initial mass, metallicity, and age by interpolating over the dust-free SED templates.

After generating the SEDs, a wavelength grid with 1168 points spanning $0.05~\mu m$ to $5.0~\mu m$ is constructed to cover the relevant \textit{JWST}/NIRCam wide filters. This grid, together with the positions and inferred smoothing lengths of the stellar particles, is provided to \texttt{SKIRT} to obtain a smoothed photon source distribution function used to isotropically launch a wavelength-dependent number of photon packets. Next, the positions, densities, and metallicities of the Voronoi gas cells are used to determine the dust distribution within the simulated galaxies, assuming a \citet{DraineLi07} dust mixture of amorphous silicate and graphitic grains with varying amounts of polycyclic aromatic hydrocarbons particles. Specifically, the distribution of dust in the interstellar medium (ISM) is determined by first computing the metal mass distribution from the metallicities of the cold ($<8000~\rm K$), star-forming ($\rm{SFR} > 0$) gas cells in the simulated galaxies. Under the assumption that dust traces metals in the interstellar medium, the metal mass distribution is then converted into a dust mass distribution using a redshift-dependent, spatially constant dust-to-metal ratio given by $0.9 \times (z/2)^{-1.92}$, calibrated using observed UV luminosity functions at $z=2-10$.

The emitted photon packets propagate through the resolved ISM, interacting stochastically with dust cells before being collected by a detector 1 Mpc from the galaxy. The recorded flux is limited to contributions from stellar particles within a physical aperture of 30 pkpc centered on the galaxy and includes IGM absorption following the models of \citet{Madau95} and \citet{Madau96}. The integrated flux is convolved with the band transmission curves using the \texttt{SEDPY}\footnote{\href{github.com/bd-j/sedpy}{github.com/bd-j/sedpy}} code to compute rest-frame magnitudes, with apparent magnitudes obtained by applying the appropriate redshift to the rest-frame flux.

This modeling procedure also includes two resolution corrections to account for (i) the coarse sampling of star formation in TNG300 and (ii) the inability to resolve birth-cloud absorption associated with recent star formation. The former is addressed by applying a resolution correction factor derived using the estimated fluxes from TNG50-1, which is the highest-resolution run in the IllustrisTNG suite. The latter is addressed by replacing the \texttt{FSPS} spectral templates for stellar particles with ages less than $10$ Myr with the \texttt{MAPPINGS-III} spectral library \citep{Groves08}, which is specifically constructed using 1D photoionization and radiative transfer calculations to model the dust attenuation of stellar light from massive young star clusters embedded within \hbox{\sc Hii} and photodissociation regions. The final data products from this procedure include dust-attenuated rest-frame and apparent magnitudes for simulated galaxies measured at nine snapshots from $z=2$ to $z=10$ for the following \textit{JWST}/NIRCam wide filters: $F070W$, $F090W$, $F115W$, $F150W$, $F200W$, $F277W$, $F356W$, and $F444W$.

We acknowledge that the dust-attenuated photometry derived from this modeling procedure is not without its limitations. Specifically, as explored in \citet{Shen20}, massive/bright galaxies in the simulation at $z\gtrsim4$ exhibit bluer UV continuum slopes than observed, suggesting that the model underestimates the amount of dust in high-redshift massive galaxies. A consequence of this underestimation is that massive galaxies in the simulation will be systematically bluer than expected, potentially causing some of them to be excluded from the \textit{JWST}/NIRCam color selections utilized in the observational literature. We discuss how this limitation impacts the conclusions drawn from our analysis in \S\ref{subsec:5.3}.

\subsection{Simulated Galaxy Sample}
\label{subsec:2.3}

The objective of this analysis is to apply \textit{JWST}/NIRCam photometric selections from the observational literature, used to identify high-redshift massive galaxy candidates, to a simulated sample of galaxies with synthetic dust-attenuated photometry. We limit our analysis to simulated galaxies at $z=5$, as this is the highest redshift snapshot in TNG300 for which we have a sufficiently large sample of galaxies with dust-attenuated photometry. Additionally, $z=5$ falls within the redshift range over which massive galaxy candidates have been identified using \textit{JWST}/NIRCam color selections from the observational literature. Thus, we construct our primary galaxy sample to include all galaxies at $z=5$ with stellar masses greater than $100$ times the baryonic mass resolution, $100 \times m_{\rm{baryon}}$, within twice the stellar half-mass radius. This resolution-driven selection yields 12,035 galaxies at $z=5$ with stellar masses in the range $\mstar \sim 10^{9.0} - 10^{11.4}~\msun$. By design this selection includes all simulated galaxies with dust-attenuated \textit{JWST}/NIRCam photometry from \citet{Vogelsberger20} and \citet{Shen20}. We exclude extremely faint galaxies with less reliable photometry by imposing an apparent magnitude cut of $m_{150W} < 29.5$, which reduces the number of galaxies in our primary sample to 11,874.

\section{Reliability of NIRCam Photometric Selections in Identifying Simulated High-Redshift Massive Galaxies}
\label{sec:NIRCam_selection_results}

\subsection{Testing Published Photometric Selection Criteria}
\label{subsec:3.1}

We compare five distinct \textit{JWST}/NIRCam selections from the observational literature (see Table~\ref{tab:NIRCam_selections}) that have been used to identify massive ($\mstar >10^{10}~\msun$) galaxy candidates at $z>3$. We note, however, that many of these selections were originally constructed to identify and characterize optically faint galaxies (OFGs) –– defined as galaxies that are simultaneously faint in optical bands and bright in the near-IR -- and were not specifically developed to identify high-redshift massive galaxies. Nevertheless, a substantial fraction of the galaxies identified by these selections were found to be massive, dusty, high-redshift galaxies that had not been identified in previous rest-frame UV surveys. An overview of each NIRCam selection is provided below: 

\begin{itemize}
    \item S1 (\citealt{PerezGonzalez23}): This selection was used to characterize a sample of $138$ OFGs identified in the \textit{JWST}/CEERS survey. The bulk of these candidates were ultimately characterized as dusty star-forming galaxies with stellar masses between $\sim 10^{9-11}~\msun$ at redshifts $2 \lesssim z \lesssim 6$. As described below, we will refer to S1 as our \emph{fiducial selection}, and we emphasize that it has previously been used to identify high-redshift galaxies in \textit{JWST} data, in contrast to the simulation-derived selections we introduce later in this work.
    \item S2 (\citealt{GomezGuijarro23}): This selection was designed to constrain the nature of 35 OFGs identified in the \textit{JWST}/CEERS survey. The majority of the OFGs (25/35) identified with this selection were characterized as high-redshift ($3 \lesssim z \lesssim 7.5$), dusty, intermediate-to-massive ($\mstar=10^{9-11}~\msun$), star-forming galaxies.
    \item S3 (\citealt{Gottumukkala24}): This selection was optimized to detect OFGs and identified $148$ massive, dusty galaxies at $3 \lesssim z \lesssim 8$ in the \textit{JWST}/CEERS survey. This selection is nearly identical to the photometric selection presented in \citet{Barrufet23}, with the primary difference being that \citet{Barrufet23} used the \textit{HST} WFC3/IR F160W filter instead of the \textit{JWST}/NIRCam F150W filter. Thus, we assume that most galaxies identified by S3 would also be identified by the selection from \citet{Barrufet23}.
    \item S4 (\citealt{Labbe23}): This selection was optimized to detect \textquote{double-break galaxies} at $7.4 \lesssim z \lesssim 9.1$, which are galaxies with SEDs exhibiting both a Lyman break at $\lambda_{\mathrm{rest}}=1216~\text{\AA}$ and Balmer break at $\lambda_{\mathrm{rest}}\sim 3600~\text{\AA}$. In practice these galaxies are undetected in the $HST$/ACS optical bands, blue in the NIRCam short-wavelength filters, and red in the NIRCam long-wavelength filters. 
    \item S5 (\citealt{Akins23}): This selection was designed to detect extremely red, dust-obscured galaxies at $6 \lesssim z \lesssim 7$ using \textit{JWST} NIRCam and MIRI\footnote{Our analysis omits the MIRI criterion from \citet{Akins23} as our simulated galaxy sample lacks the requisite synthetic mid-infrared photometry. However, we note that the inclusion of the massive galaxy candidates identified in \citet{Akins23} is insensitive to the MIRI selection criterion, as it was primarily designed to reject objects with red mid-IR SEDs (i.e., dust-obscured AGNs).} imaging from the COSMOS-Web, CEERS, and PRIMER surveys, and was used to identify two massive galaxy candidates ($\mstar > 10^{10}~\msun$) at $z\sim8.5$ and $z\sim7.6$.
\end{itemize}

\begin{figure}
\centering
\hspace*{-0.15in}
\includegraphics[width=3.5in]{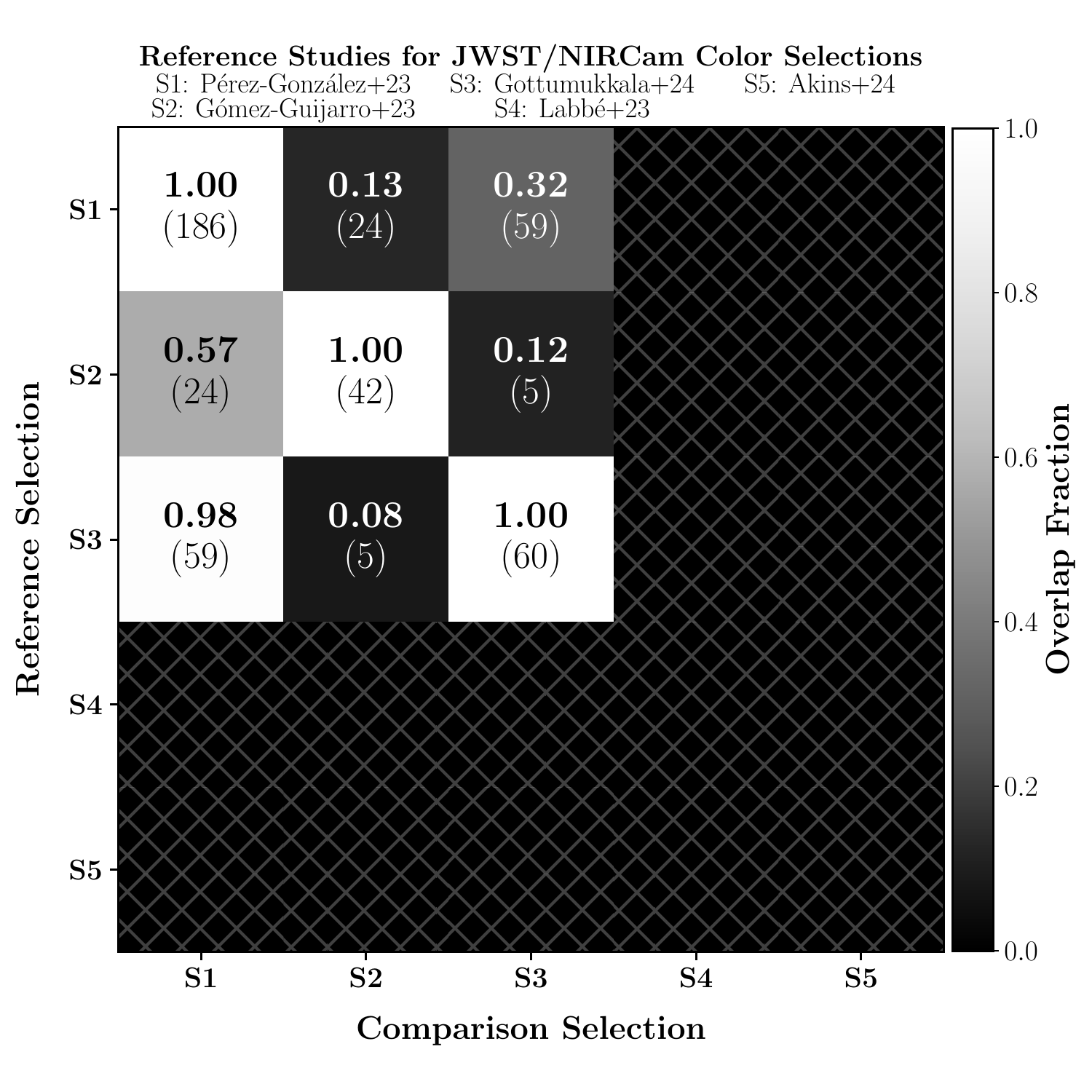}
\caption{Matrix showing the fractional overlap between simulated galaxy populations at $z=5$ selected using the five NIRCam photometric criteria described in Table~\ref{tab:NIRCam_selections}. Each cell gives the fraction and number of galaxies selected by the row criterion (reference selection) that are also selected by the column criterion (comparison selection). Cells marked with cross hatching indicate that no simulated galaxies were identified by the reference selection.}
\label{fig:fig1}
\end{figure}

Fig.~\ref{fig:fig1} depicts the overlap in the five NIRCam selections listed in Table~\ref{tab:NIRCam_selections} to the simulated galaxies used here at $z=5$. This is displayed as a matrix, color-coded by the fractional overlap between galaxies selected in one reference sample (rows) and those in a comparison sample (columns). Notably, the selections presented in \citet{Labbe23} and \citet{Akins23} (S4 and S5, respectively) yield zero galaxies in the TNG300 simulation. While not included in this analysis, we also find that the NIRCam selection presented in \citet{Gentile24}, which is similar to that of \citet{Akins23} but adopts a less restrictive color cut (i.e., $m_{\rm{277W}} - m_{\rm{444W}} > 1.5$), also yields zero galaxies. We acknowledge that S4 and S5 were optimized to identify massive galaxy candidates at redshifts higher than those explored in Fig.~\ref{fig:fig1}; however, even when applied to simulated galaxy populations at $z>5$ they continue to yield non-detections. We discuss possible explanations for these non-detections in \S\ref{subsec:5.3}.

Conversely, we find that a non-zero number of simulated galaxies satisfy the first three NIRCam selections. The selection that yields the largest number of galaxies is S1, with $186$ galaxies in total, followed by S3 and S2, which yield $60$ and $42$ NIRCam-selected galaxies, respectively. We also find that S1 recovers 24/42 (57\%) of the galaxies selected by S2 and 59/60 (98\%) of those selected by S3. The 18 galaxies selected by S2 but not S1 have $m_{\rm{150W}}-m_{\rm{356W}}$ colors between $\sim1.36$ and $1.49$ mag, placing them just below the S1 color cut of $m_{\rm{150W}}-m_{\rm{356W}}>1.5$. This arises because S2 uses two magnitude cuts (as opposed to a magnitude and color cut) that create a wedge in $m_{\rm{150W}}-m_{\rm{356W}}$ versus $m_{\rm{356W}}$ color--magnitude space with a vertex at approximately $(m_{\rm{356W}},m_{\rm{150W}}-m_{\rm{356W}})=(25.2, 1.3)$, thereby including galaxies slightly bluer than allowed by S1. In contrast, S3 also uses a color and magnitude cut similar to S1, albeit slightly more restrictive, limiting the selection to galaxies with $m_{\rm{150W}}-m_{\rm{356W}}>1.8$ mag. Given that S1 encompasses the majority of galaxies selected by both S2 and S3 at $z=5$, we adopt it as the fiducial selection for the remainder of our analysis.  

\begin{figure*}
\centering
\hspace*{-0.25in}
\includegraphics[width=5.75in]{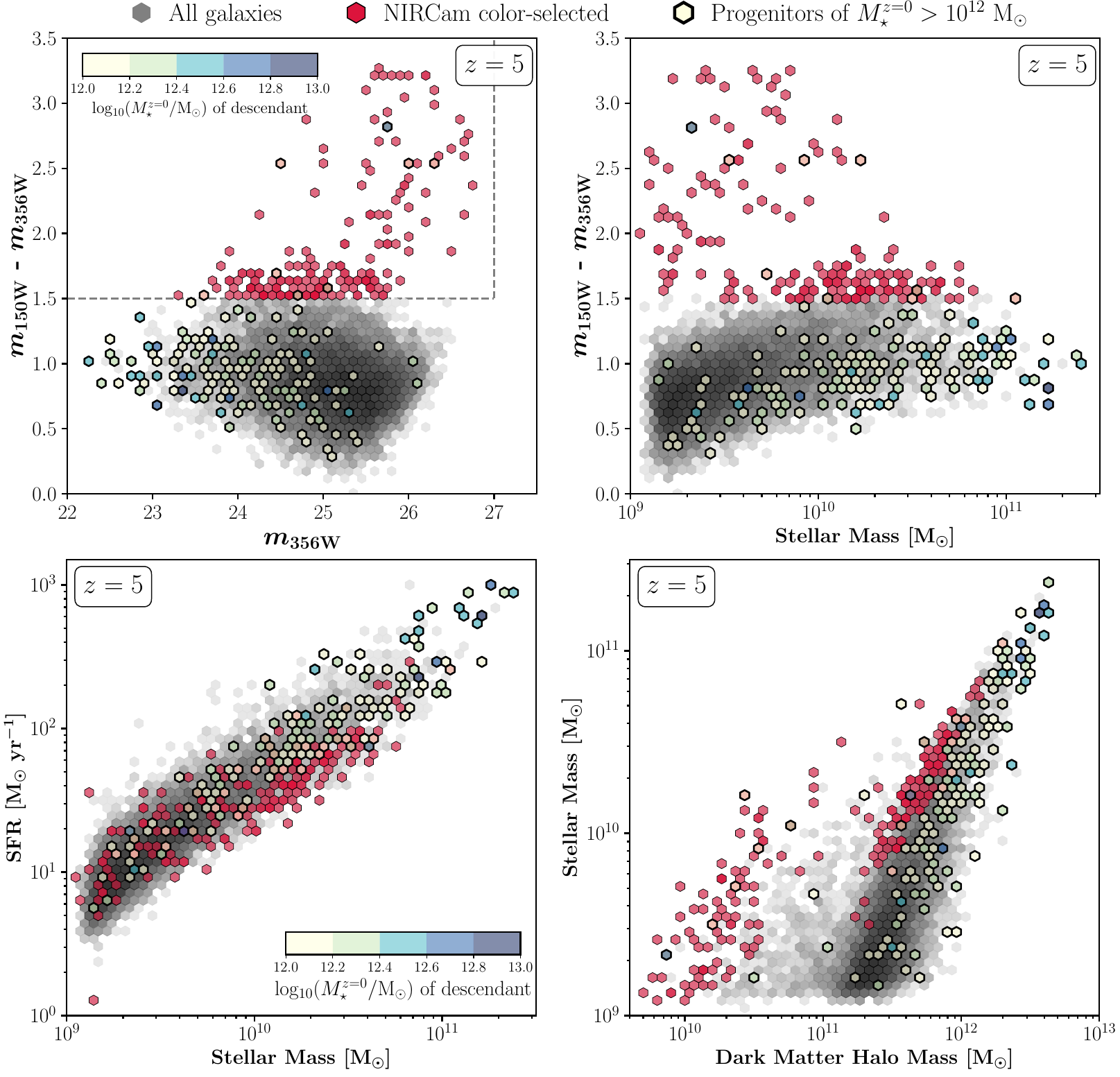}
\caption{Panel plot showing color versus magnitude (top left), color versus stellar mass (top right), star formation rate versus stellar mass (bottom left), and stellar mass versus halo mass (bottom right) for simulated galaxies at $z=5$. The full galaxy population is represented by black and gray hexagonal bins. The subpopulation of NIRCam-selected galaxies, identified using the criteria of \citet{PerezGonzalez23}, is shown by red hexagonal bins, while the progenitors of galaxies with stellar masses exceeding $10^{12}~\msun$ at $z=0$ are indicated by hexagonal bins with thick black edges and colored according to their $z=0$ stellar mass, as shown by the color bar. The primary takeaways are: (i) the NIRCam color selection is inefficient at identifying the progenitors of the most massive galaxies at $z=0$; and (ii) the NIRCam-selected galaxies exhibit lower star formation rates and reside in less massive dark matter (sub)halos than the progenitors of the most massive galaxies at $z=0$.}
\label{fig:fig2}
\end{figure*}

\begin{figure*}
\centering
\hspace*{-0.25in}
\includegraphics[width=6.5in]{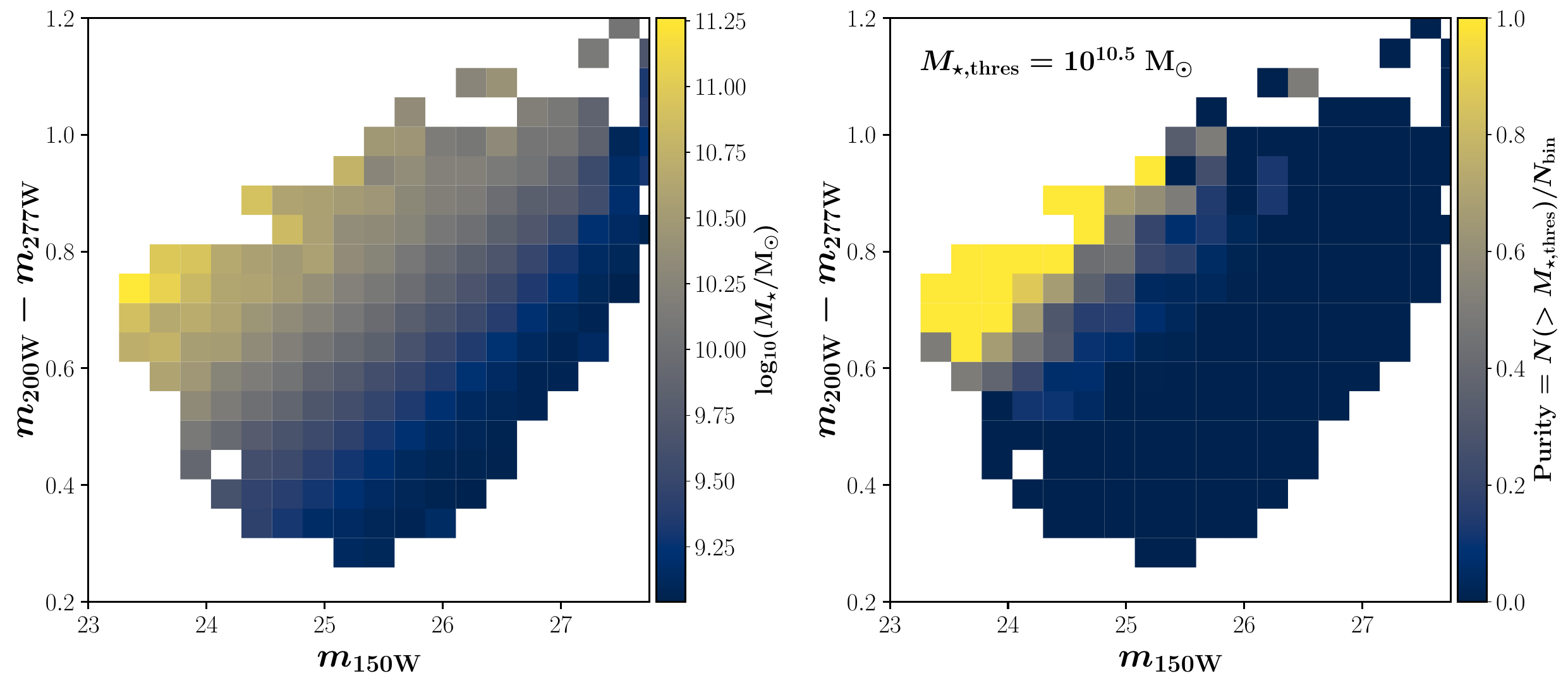}
\caption{\textit{JWST}/NIRCam color–magnitude diagrams at $z=5$ constructed from the $\mathrm{F200W}$, $\mathrm{F277W}$, and $\mathrm{F150W}$ filters. The left (right) panel is color-coded by mean stellar mass (purity). We define purity as the fraction of TNG300 galaxies in each color–magnitude bin with $\mstar > 10^{10.5}~\msun$. In this parameter space, the most massive simulated galaxies are efficiently selected with high purity using $m_{\mathrm{150W}} < 25.3$ mag and $(m_{\mathrm{200W}} - m_{\mathrm{{277W}}}) > 0.21\times m_{\mathrm{150W}} - 4.4$ mag.}
\label{fig:fig3}
\end{figure*}

\subsection{Properties of NIRCam-selected Galaxies}
\label{subsec:3.2}

The top-left panel of Fig.~\ref{fig:fig2} shows the color-magnitude diagram ($m_{\rm{150W}} - m_{\rm{356W}}$ versus $m_{\rm{356W}}$) for our simulated galaxy population at $z=5$. The full population is represented by the black and gray hexagons, and the subpopulation that satisfies the fiducial S1 selection is represented by the red hexagons. Additionally we highlight the progenitors of galaxies with stellar masses exceeding $10^{12}~\msun$ at $z=0$ using hexagons with thick black edges, color-coded according to the stellar mass of their present-day descendant. We find that only 8 out of the 186 (4\%) NIRCam-selected galaxies are progenitors of galaxies with stellar masses above $10^{12}~\msun$ at $z=0$ in the TNG300 simulation, and that the bulk of the progenitors of present-day massive galaxies are bluer in $m_{\rm{150W}} - m_{\rm{356W}}$ than the NIRCam-selected sample. 

The top-right panel of Fig.~\ref{fig:fig2} shows $m_{\rm{150W}} - m_{\rm{356W}}$ versus stellar mass for the TNG simulated galaxy population at $z=5$. More than half (97 out of 186) of the NIRCam-selected galaxies (using the fiducial S1 selection) have stellar masses exceeding $10^{10}~\msun$ at $z=5$. However, only 1 out of the 17 galaxies with stellar masses exceeding $10^{11}~\msun$ are included in the fiducial NIRCam selection at $z=5$. We find that galaxies with $\mstar > 10^{11}~\msun$ have an average $m_{\rm{150W}} - m_{\rm{356W}}$ color of $\sim 1.0$ mag, which is $\sim 0.5$ mag bluer than the single NIRCam-selected galaxy in this stellar mass range. Our finding that the most massive galaxies at $z=5$ in the TNG300 simulation are largely absent from the fiducial NIRCam selection is consistent with recent studies suggesting that the SED-derived stellar masses of NIRCam-selected high-redshift massive galaxy candidates may be overestimated \citep[e.g.,][]{Chworowsky24, WangTao25}. Nevertheless, in \S\ref{subsec:5.3} we discuss the limitations of the procedure used to estimate synthetic dust-attenuated photometry for simulated galaxies in TNG300 and explain why the abundance of massive NIRCam-selected galaxies inferred from this analysis should be regarded as a lower limit.

The bottom-left panel of Fig.~\ref{fig:fig2} depicts SFR versus stellar mass for our simulated galaxy population at $z=5$. At fixed stellar mass, we find that the NIRCam-selected galaxies exhibit lower SFRs compared to the progenitors of galaxies with stellar masses exceeding $10^{12}~\msun$ at $z=0$. Lastly, the bottom-right panel of Fig.~\ref{fig:fig2} depicts the stellar-to-halo mass relation for our simulated galaxy population at $z=5$. At fixed stellar mass the NIRCam-selected galaxies preferentially occupy less massive dark matter (sub)halos compared to the progenitors of the most massive galaxies at $z=0$. Additionally, we find that the NIRCam-selected galaxies fall into two distinct subpopulations split above and below $\mhalo \sim 10^{11}~\msun$. We find that these two subpopulations are also present when selecting galaxies using S2 instead of our fiducial NIRCam selection, S1. This is expected, as S2 is largely a subset of S1. In contrast, the S3 selection, which is a subset of the faintest and reddest galaxies selected by S1, consists exclusively of galaxies from the subpopulation below $\mhalo \sim 10^{11}~\msun$. On the surface, this seems to suggest that S3 struggles to select galaxies more massive than $\mstar \gt 10^{10}~\msun$. However, the performance of S3 at identifying massive galaxies varies considerably with redshift, with this selection capturing a larger fraction of galaxies with $\mstar > 10^{10}~\msun$ at $z\lt5$. This is consistent with \citet{Gottumukkala24}, who found that massive galaxies identified by this selection predominantly reside at $z<5$.

We analyze the star formation and orbital histories of the galaxies in these two subpopulations to better understand their origins. The subpopulation with $\mhalo \gtrsim 10^{11}~\msun$ primarily consists of central galaxies that lie on the locus of the TNG300 stellar-to-halo mass relation at $z=5$. We find that these galaxies are preferentially identified by the fiducial NIRCam color selection because they experienced previous episodes of prodigious star formation, which enriched their interstellar medium and led to elevated gas-phase metallicities. This, in turn, results in higher dust content and redder $m_{\rm{150W}} - m_{\rm{356W}}$ colors at $z=5$.

The subpopulation with $\mhalo \lesssim 10^{11}~\msun$ consists of galaxies with severely stripped dark matter halos due to prior tidal interactions with more massive hosts; we find that these galaxies have lost on average $86\%$ of their peak halo mass, as measured prior to $z=5$. This tidal stripping is accompanied by a steep decline in the overall gas fraction, as well as increases in both the star formation rate and gas-phase metallicity. Following from the assumption of a spatially constant dust-to-metal ratio, these galaxies exhibit relatively high dust content and correspondingly red $m_{\rm{150W}} - m_{\rm{356W}}$ colors at $z=5$, driving their inclusion in our fiducial NIRCam selection. In summary, we find that the \textit{JWST}/NIRCam selections explored in this analysis preferentially identify galaxies in TNG300 with elevated gas-phase metallicities resulting from recent bursts of star formation, regardless of whether the bursts are driven by wet mergers, inflow of gas from the IGM, or tidal interactions with a more massive host.

\subsection{Improved NIRCam Color Selection Criteria}
\label{subsec:3.3}

Accurate measurements of the number density of the most massive galaxies in the high-$z$ universe provide key constraints on models of galaxy formation and evolution. In the current era of large near-infrared imaging surveys, it is therefore imperative to develop photometric selections that identify high-quality high-redshift massive galaxy candidates for spectroscopic confirmation. In the previous subsection we found that while the fiducial NIRCam color selection is fairly effective at identifying high-redshift, massive galaxies with $\mstar = 10^{10-11}~\msun$, it is far less effective at identifying the most massive high-redshift galaxies with stellar masses above $10^{11}~\msun$. In this subsection we use the sample of simulated galaxies with synthetic dust-attenuated NIRCam photometry to define novel \textit{JWST}/NIRCam color-magnitude selections that can be used to efficiently identify the most massive galaxies at $z\sim5$. 

We begin by computing all possible color-magnitude permutations constructed from the following dust-attenuated NIRCam apparent magnitudes: $m_{\rm{150W}}$, $m_{\rm{200W}}$, $m_{\rm{277W}}$, $m_{\rm{356W}}$, and $m_{\rm{444W}}$. For each unique combination, we iteratively apply wedge-based color-magnitude selections and evaluate their associated purity and completeness. To achieve this we set a minimum stellar mass threshold of $M_{\star,\mathrm{thres}} = 10^{10.5}~\msun$ and define \textquote{purity} as the fraction of galaxies identified in a given color-magnitude selection with $M_\star \geq M_{\star,\mathrm{thres}}$. Similarly, we define \textquote{completeness} as the number of galaxies above $M_{\star,\mathrm{thres}}$ in a given color-magnitude selection divided by the total number of galaxies above $M_{\star,\mathrm{thres}}$ in the full sample. We then filter our results to identify the color-magnitude selection that maximizes purity while maintaining a baseline completeness above $20\%$, as well as the selection that maximizes completeness while maintaining a baseline purity above $20\%$.

\begin{figure}
\centering
\hspace*{-0.25in}
\includegraphics[width=3.5in]{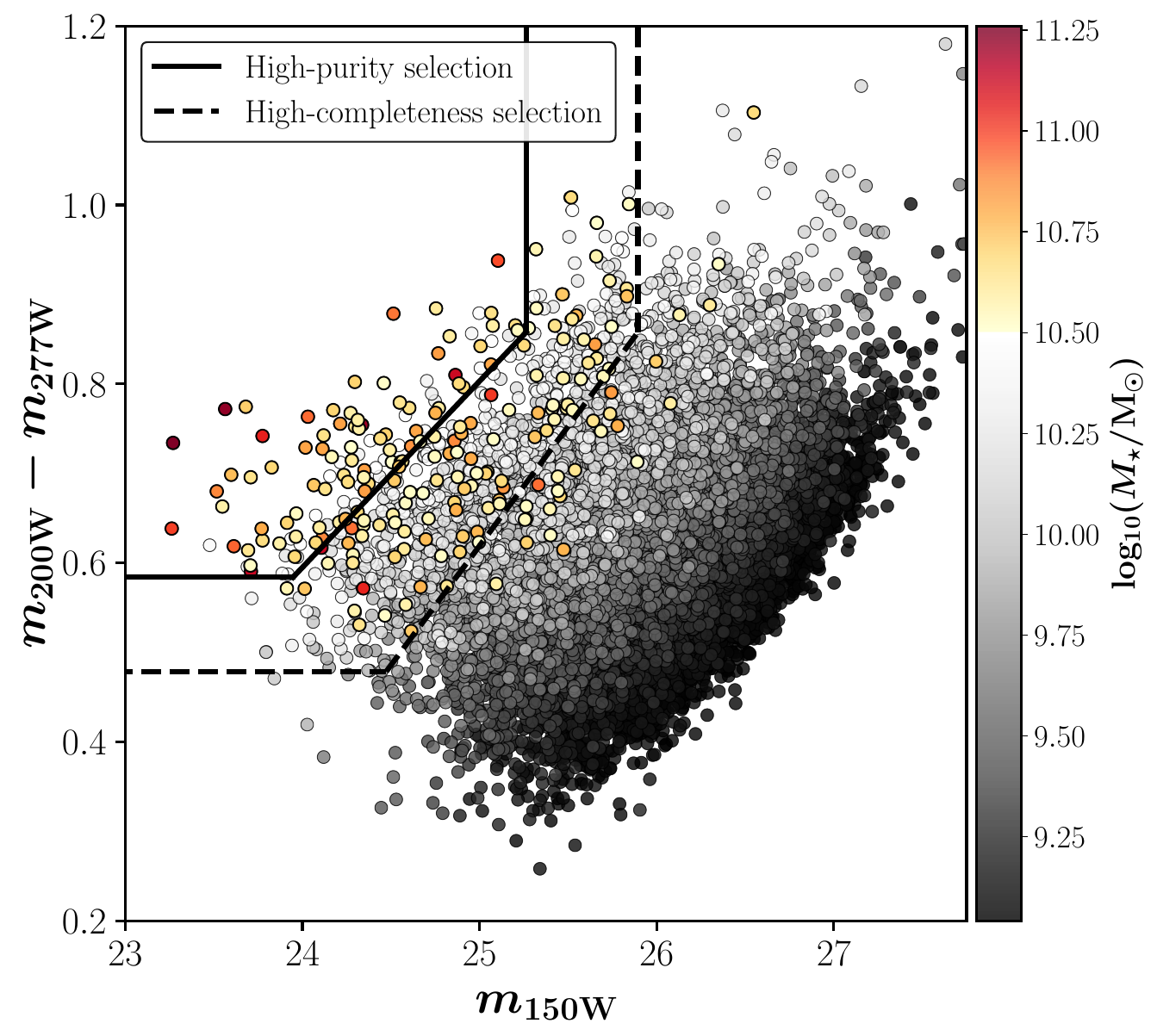}
\caption{Similar to Fig. 3, with black and white (colored) circles representing TNG300 galaxies with stellar masses below (above) $\mstar = 10^{10.5}~\msun$. The solid (dashed) wedge represents the region in which simulated galaxies more massive than $\mstar = 10^{10.5}~\msun$ at $z=5$ are selected with high purity (high completeness). Specifically, the solid wedge has a purity of $0.70$ and a completeness of $0.39$, while the dashed wedge has a purity and completeness of $0.20$ and $0.88$, respectively.}
\label{fig:fig4}
\end{figure}

From this analysis we find that the combination of the $m_{\rm{200W}}$, $m_{\rm{277W}}$, and $m_{\rm{150W}}$ dust-attenuated apparent magnitudes is most effective at segregating galaxies by stellar mass at $z=5$ in TNG300. This is illustrated by the prominent stellar mass gradient in the $m_{\rm{200W}} - m_{\rm{277W}}$ versus $m_{\rm{150W}}$ color-magnitude diagram (left-hand panel of Fig.~\ref{fig:fig3}), which smoothly transitions from massive galaxies in the upper left to low-mass galaxies in the lower right. Similarly, color-coding this diagram by purity (right-hand panel of Fig.~\ref{fig:fig3}) reveals a region in the upper left where galaxies more massive than $M_{\star,\mathrm{thres}} = 10^{10.5}~\msun$ at $z=5$ can be efficiently selected with high purity. We use the $m_{\rm{200W}}$, $m_{\rm{277W}}$, and $m_{\rm{150W}}$ magnitudes to define both a high-purity and a high-completeness selection, illustrated by the solid and dashed wedges in Fig.~\ref{fig:fig4}. These simulation-motivated selections, denoted S6 and S7, are defined as follows:
\begin{equation*}
\label{eqn:color_selection}
\begin{aligned}
\mathrm{S6}=\enspace  
& \begin{cases}
(m_{\rm{200W}} - m_{\rm{277W}}) > 0.58 \\
(m_{\rm{200W}} - m_{\rm{277W}}) > 0.21\times m_{\rm{150W}} - 4.4 \\
m_{\rm{150W}} < 25.3
\end{cases} \\
\mathrm{S7}=\enspace  
& \begin{cases}
(m_{\rm{200W}} - m_{\rm{277W}}) > 0.48 \\
(m_{\rm{200W}} - m_{\rm{277W}}) > 0.27\times m_{\rm{150W}} - 6.0\\
m_{\rm{150W}} < 25.9. 
\end{cases}
\end{aligned}
\end{equation*}

\begin{figure*}
\centering
\hspace*{-0.25in}
\includegraphics[width=7.5in]{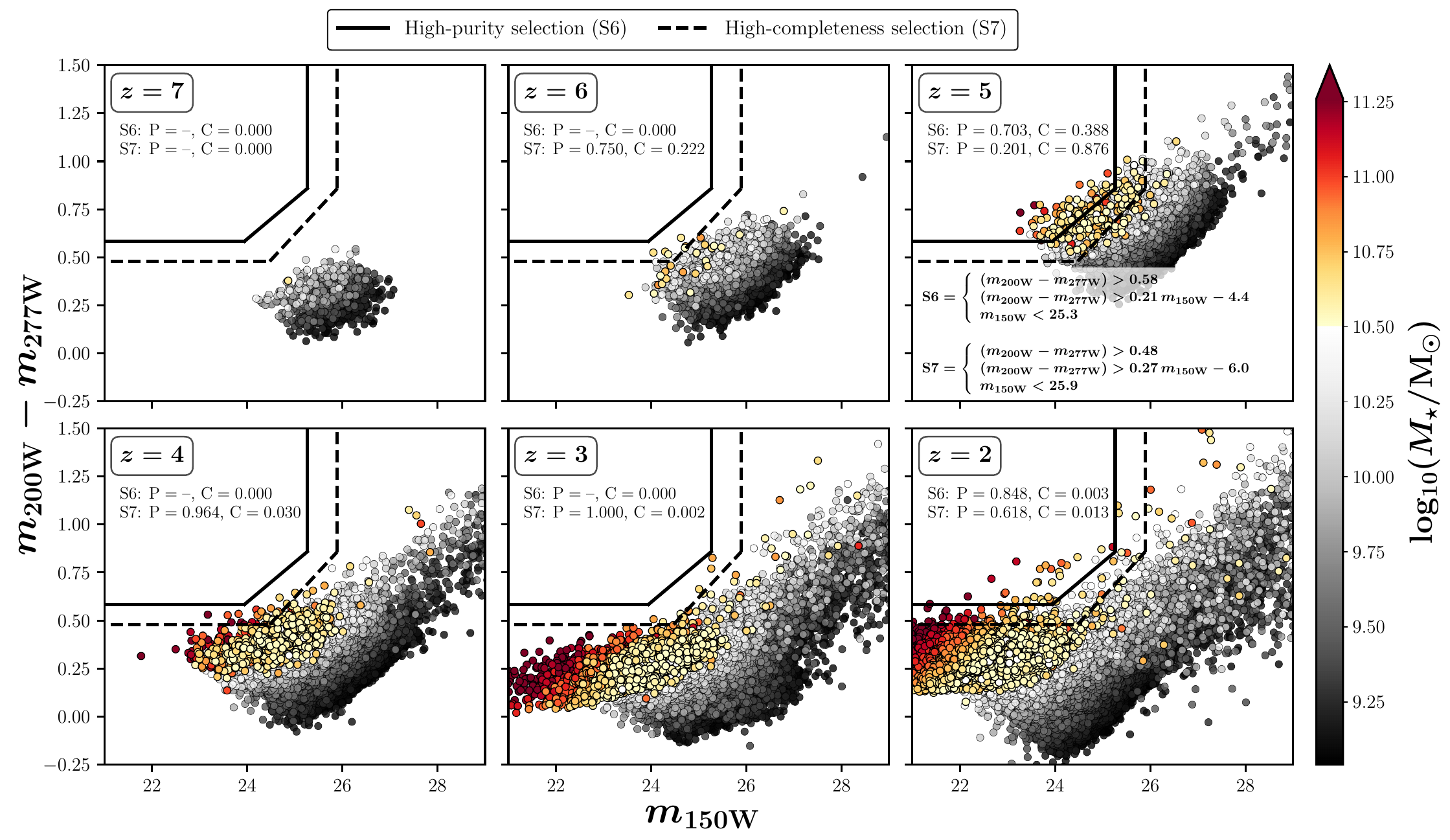}
\caption{Similar to Fig.~\ref{fig:fig4}, but with the addition of TNG300 galaxies at $z=7,~6,~4,~3,~$ and $2$ with $\mstar>10^{9.0}~\msun$. The purity and completeness of the NIRCam selections at each redshift snapshot are provided in top-left corner of each plot. The simulation-derived S6 selection, which was optimized to identify massive galaxies ($\mstar > 10^{10.5}~\msun$) at $z=5$ with high purity, does not identify any galaxies at $z=3$, $4$, $6$, or $7$, but selects massive galaxies with high purity at $z=2$. The simulation-derived S7 selection, which was optimized to identify massive galaxies at $z=5$ with high completeness, does not identify any galaxies at $z=7$. However, this selection does identify massive galaxies with high purity at $z=2,~3,~4$ and $6$, albeit with very low completeness.}
\label{fig:fig5}
\end{figure*}

\begin{figure*}
\centering
\hspace*{-0.25in}
\includegraphics[width=5.5in]{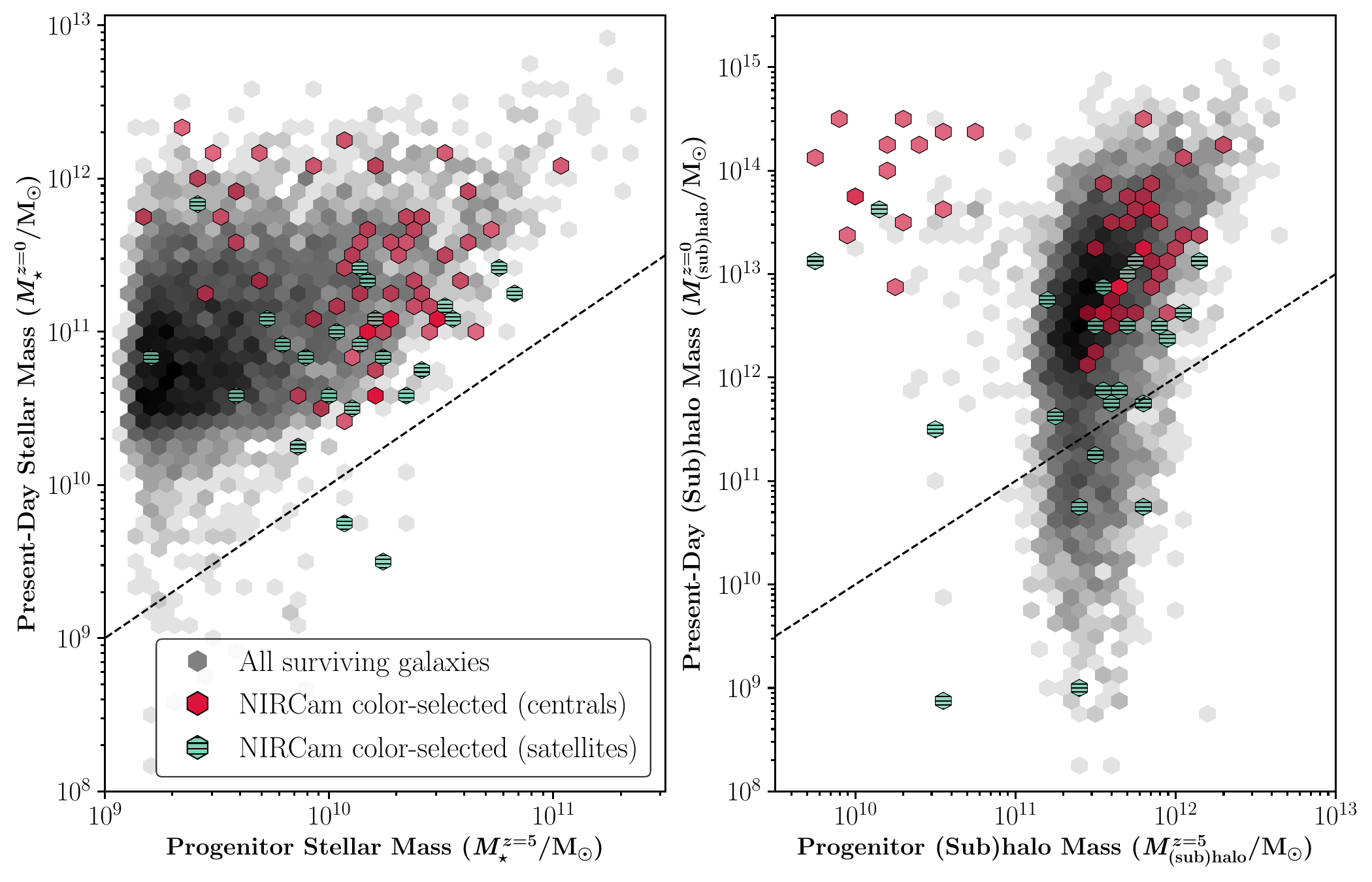}
\caption{Comparison of the present-day and descendant stellar and (sub)halo masses (left and right panels, respectively) for the subsample of TNG300 galaxies that are selected at $z=5$ by our fiducial criteria and survive to $z=0$ without being disrupted. The black and gray hexagonal bins represent the full galaxy population, while the red (blue-hatched) hexagonal bins represent NIRCam-selected galaxies that are centrals (satellites) at $z=0$. We find that the descendants of the NIRCam-selected galaxies are generally neither the most massive galaxies nor do they occupy the most massive dark matter halos at $z=0$.}
\label{fig:fig6}
\end{figure*}

The \emph{high-purity} selection (S6) yields a purity of 0.70 and a completeness of 0.39, while the \emph{high-completeness} selection (S7) yields a purity of 0.20 and a completeness of 0.88. While these NIRCam color selections are derived using a data-driven approach designed to optimize the identification of massive galaxies at $z=5$, their effectiveness stems from the fact that they straddle the Balmer break ($\lambda_{\rm{rest}}=3646~$\AA) at this redshift. Specifically, at $z=5$ the Balmer break is redshifted to $\sim2.19~\mu$m, which falls directly between the $F200W$ and $F277W$ filters. As such, the $m_{\rm{200W}} - m_{\rm{277W}}$ color is sensitive to the strength of the Balmer break, which in itself depends on the age and star formation history of a galaxy. We note that analogous color selections designed to probe the strength of the Balmer discontinuity have been used in pre-\textit{JWST} searches for massive galaxies at $z > 3$ \citep[e.g.,][]{Wiklind08, Nayyeri14, Mawatari16, AlcaldePampliega19, Mawatari20, Shahidi20}, but here we have updated this selection for \textit{JWST}/NIRCam filters. Unlike the NIRCam selections presented in Table~\ref{tab:NIRCam_selections}, which are defined by a single color cut and a magnitude threshold, our selections employ a magnitude-dependent color cut to identify massive galaxies at $z\sim5$. As such, our updated selections naturally exclude the subpopulation of heavily stripped, low-mass galaxies picked up in the fiducial S1 NIRCam selection. We emphasize, however, that these photometric selections are specifically optimized to detect massive galaxies in TNG300 at $z=5$. In the following subsection, we evaluate the purity and completeness of these selections when applied to simulated galaxies at lower and higher redshifts.

\subsection{Performance of Simulation-Derived \textit{JWST}/NIRCam Selections Across Redshift Snapshots}
\label{subsec:3.4}

n Fig.~\ref{fig:fig5} we explore how the simulation-derived NIRCam selections presented in \S\ref{eqn:color_selection} perform when applied to redshift snapshots other than $z=5$. We find that the high-purity selection (S6) does not identify any galaxies in TNG300 at $z=3,~4,~6,~$ or $7$. However, this selection does capture a small population of galaxies at $z=2$. Nearly all of these galaxies exceed the stellar mass threshold of $\mstar=10^{10.5}~\msun$, resulting in a relatively high purity of $0.85$. Nevertheless, this selection only captures a small fraction of the total massive galaxy population at $z=2$, yielding a completeness much less than $1\%$.

Conversely, the high-completeness selection (S7) identifies eight galaxies at $z=6$ and zero galaxies at $z=7$. Six of the eight galaxies identified at $z=6$ are above the stellar mass threshold, resulting in a relatively high purity ($0.75$) and low completeness ($0.22$) at a slightly higher redshift than the $z=5$ snapshot at which the selection was optimized. At lower redshifts, S7 also identifies massive galaxies in TNG300 with high purity but extremely low completeness ($\ll 1\%$).

Together, these findings demonstrate that the NIRCam selections developed in this analysis, which are specifically optimized at $z=5$, can identify massive galaxies in TNG300 with relatively high purity and low completeness across a broad range of redshifts. Given that the color-magnitude space probed by these selections is largely occupied by massive galaxies in the simulation, they may prove useful for observers seeking to select massive galaxies between $2 \lesssim z \lesssim6$ in photometric surveys with relatively high purity. 

We also investigate whether modest modifications to these selections can improve their overall performance at lower and higher redshifts. At $z<5$, we find that adopting a bluer color threshold for S7, $(m_{\mathrm{200W}} - m_{\mathrm{277W}}) > 0.2$, substantially increases completeness with only a modest reduction in purity. Specifically, the completeness increases from $\sim0.01$ to $\sim0.94$ at $z=2$, and from $<0.03$ to $\sim0.3$ at $z=3$ and $4$, with a corresponding purity drop of less than $0.20$ at $z<5$

At $z>5$, modifying the color threshold alone is insufficient to improve these selections because galaxies are systematically bluer and fainter relative to those at $z=5$. Therefore, to improve the S7 selection, we adopt a bluer color threshold of $(m_{\mathrm{200W}} - m_{\mathrm{277W}}) > 0.2$ and shift the $y$-intercept from $-6.0$ to $-6.15$ at $z=6$ and $-6.20$ at $z=7$. Moreover, because simulated galaxies in TNG300 with $\mstar>10^{10.5}~\msun$ are less common at $z>5$, we evaluate the purity and completeness of these modified selections using a lower stellar mass threshold of $\mstar>10^{10}~\msun$. At $z=6$, these adjustments yield a purity and completeness of $0.74$ and $0.39$, respectively. Similarly, at $z=7$, the purity and completeness are $0.71$ and $0.13$, respectively.

\begin{figure*}
\centering
\hspace*{-0.3in}
\vspace*{0.8in}
\includegraphics[width=4.75in]{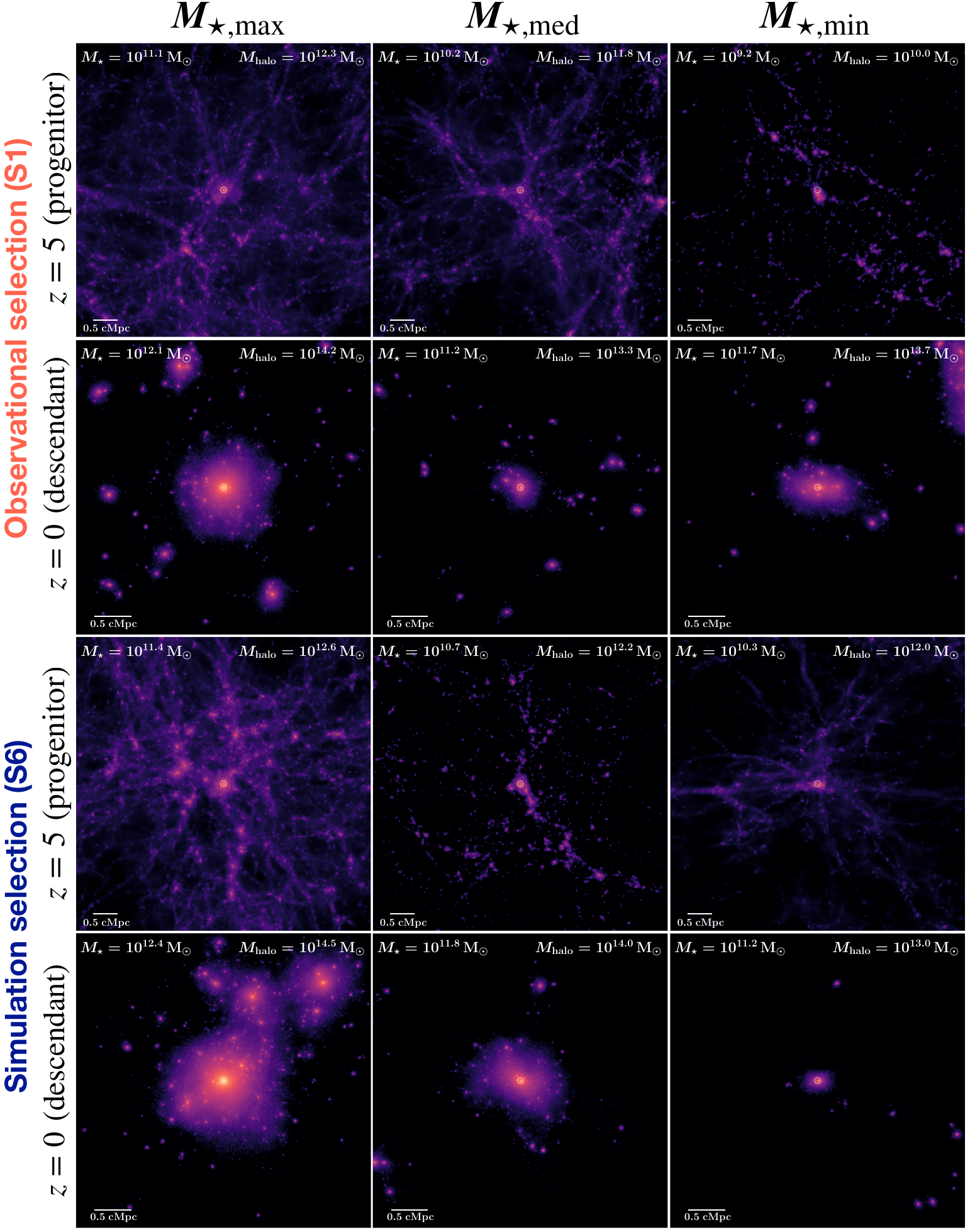}
\vspace{-0.75in}
\caption{Surface density of dark matter centered on the most massive, median-mass, and least massive galaxies (left, middle, and right columns, respectively) identified using our fiducial NIRCam selection (S1, top two rows) and high-purity selection (S6, bottom two rows). The first and third rows show the large-scale environments of these galaxies at $z=5$, while the second and fourth rows show how these environments evolved by $z=0$. The dynamic range of the dark matter surface density is fixed at each redshift to enable direct visual comparison between systems. Galaxies identified by the high-purity selection are typically more massive at both $z=5$ and $z=0$ than their counterparts in the fiducial selection. However, the descendant of the least massive galaxy in the fiducial selection is more massive than the descendant of the least massive galaxy in the high-purity selection, highlighting that relatively low-mass galaxies at $z=5$ can outgrow their more massive coeval counterparts by $z=0$.}
\label{fig:fig7}
\end{figure*}

\section{Connecting High-Redshift Massive Galaxies to Present-Day Descendants}
\label{sec:evolution_results}  

\subsection{Properties of the Present-Day Descendants of NIRCam-selected Galaxies}
\label{subsec:4.1}

In this section we address the central question motivating this analysis: what becomes of \textit{JWST}/NIRCam-selected high-redshift massive galaxies? This is visualized in Fig.~\ref{fig:fig6}, which compares the stellar and (sub)halo masses of simulated galaxies at $z=5$ with those of their descendants at $z=0$. This comparison is performed by following the main descendant branches of the TNG300 \texttt{SUBLINK} merger trees and is restricted to galaxies selected at $z=5$ that survive to $z=0$ without being tidally disrupted. We find that 5,730 out of 11,874 galaxies in the initial sample (79 out of 186 in the fiducial S1 NIRCam-selected subsample) satisfy this condition.

We first investigate whether the most massive galaxies identified by the fiducial S1 selection at $z=5$ remain the most massive within this subpopulation at $z=0$. This is shown in the left-hand panel of Fig.~\ref{fig:fig6}, where the black and gray hexagons represent the surviving galaxy population, and the red (blue-hatched) hexagons represent NIRCam-selected galaxies that are centrals (satellites) at $z=0$. We find that the stellar masses of the galaxies identified in the S1 selection at $z=5$ are not strongly correlated with the stellar masses of their descendants at $z=0$, indicating that the most massive NIRCam-selected galaxies at $z=5$ do not necessarily evolve into the most massive galaxies in this subsample at $z=0$. For example, the most massive descendant of a NIRCam-selected galaxy has a stellar mass of only $\mstar \sim 2\times10^{9}~\msun$ at $z=5$. Additionally, only eight descendants of the surviving NIRCam-selected galaxies have $z=0$ stellar masses exceeding $10^{12}~\msun$, and half of these galaxies were relatively low-mass at $z=5$ with stellar masses below $10^{10}~\msun$. Nevertheless, the descendants of NIRCam-selected galaxies are generally more massive than the surviving galaxy population as a whole, with a median stellar mass of $10^{11.20}~\msun$ compared to $10^{10.95}~\msun$. These results indicate that while the fiducial NIRCam selection adopted in this analysis successfully identifies progenitors of relatively massive galaxies at $z=0$, it is not particularly effective at identifying progenitors of the most massive galaxies at $z=0$.

We extend this analysis in the right-hand panel of Fig.~\ref{fig:fig6} by comparing the dark matter (sub)halo masses of simulated galaxies selected at $z=5$ with those of their surviving descendants at $z=0$. As in the stellar mass analysis, we find that the surviving descendants of galaxies identified in the fiducial S1 selection at $z=5$ do not typically reside in the most massive dark matter (sub)halos at $z=0$, nor is there a significant correlation between their (sub)halo masses at $z=5$ and $z=0$. Nevertheless, the descendants of NIRCam-selected galaxies inhabit more massive (sub)halos than the surviving galaxy population overall, with median (sub)halo masses of $10^{13.12}$ and $10^{12.66}~\msun$, respectively. Overall, these results mirror those for stellar mass in that, although NIRCam-selected galaxies preferentially evolve into galaxies residing in relatively more massive (sub)halos, they do not preferentially become the central galaxies of the most massive halos at $z=0$.

While not the primary focus of this analysis, we also find that six of the surviving NIRCam-selected galaxies reside in \emph{less massive} dark matter (sub)halos than their $z=5$ progenitors (i.e., the blue-hatched hexagons below the dashed one-to-one line in the right-hand panel of Fig.~\ref{fig:fig6}). As shown in the left-hand panel of Fig.~\ref{fig:fig6}, two of these galaxies also have lower stellar masses at $z=0$ than at $z=5$. All six of these galaxies are satellites, suggesting that their extreme mass loss is likely the result of prior tidal or merger-induced stripping. 

Conversely, we find that a few of the NIRCam-selected galaxies from the subpopulation at $z=5$ with stripped dark matter halos (see the bottom-right panel of Fig.~\ref{fig:fig2}) survive to $z=0$, where they become the most massive galaxies in group- and cluster-scale dark matter halos. This suggests that these galaxies underwent at least one major merger (or multiple minor mergers) between $z=5$ and $z=0$ that substantially increased their stellar mass. Overall, these results reinforce the conclusion that the most massive high-redshift galaxies identified by the fiducial NIRCam selection are not typically progenitors of the most massive galaxies at $z=0$. 

\subsection{The Large-Scale Environments of NIRCam-Selected Galaxies}
\label{subsec:4.2}

In the previous section we analyzed the baryonic and dark matter properties of the descendants of galaxies identified by the fiducial S1 selection, finding that the most massive galaxies in this subsample at $z=5$ do not necessarily evolve into the most massive galaxies at $z=0$. Here, we extend this analysis by investigating whether differences in the large-scale environments of galaxies identified by the fiducial S1 and high-purity S6 NIRCam selections provide additional insight into their subsequent evolutionary pathways. 

In Fig.~\ref{fig:fig7} we visualize the large-scale cosmic web environments surrounding the most, median, and least massive galaxies in the TNG300 simulation ranked by stellar mass (left, middle, and right columns, respectively) drawn from the fiducial NIRCam selection (top two rows) and high-purity selection (bottom two rows). The first and third rows (second and fourth rows) show the dark matter surface density within projected regions of size $6 \times 6$ cMpc ($4 \times 4$ cMpc), integrated along a line-of-sight depth of 10 cMpc, centered on the NIRCam-selected galaxies at $z=5$ (their descendants at $z=0$). The dynamic range of the dark matter surface density is fixed at each redshift to enable direct visual comparisons between systems, such that brighter regions correspond to higher density.

The most massive galaxies identified in both the fiducial and high-purity selections would be considered ultra-massive at $z=5$, with stellar masses exceeding $10^{11}~\msun$. Similarly, their descendants are even more massive, exceeding $10^{12}~\msun$ at $z=0$. These galaxies reside at the intersection of multiple filaments within protoclusters at $z=5$, while their descendants are the most massive galaxies in galaxy clusters at $z=0$.

While the most massive galaxies in the two selections are fairly similar, the median-mass galaxies, which better represent typical systems, are far more distinct. The median-mass galaxy identified in the fiducial selection is $\sim0.6$ dex less massive at $z=5$ than its counterpart in the high-purity selection. Although both galaxies reside within cosmic filaments at $z=5$, the median-mass galaxy in the high-purity selection occupies a noticeably denser and more evolved region, as indicated by brighter colors and more concentrated dark matter halos. In addition, the descendant of the median-mass galaxy in the high-purity selection resides in a cluster-mass halo at $z=0$, whereas the fiducial-selected counterpart resides in a group-mass halo. Although both systems are the dominant galaxies in their local environments at $z=0$, the descendant of the high-purity selected galaxy is $\sim0.5$ dex more massive.

The starkest difference between these selections emerges when comparing their least massive galaxies. The least massive galaxy identified in the fiducial selection is more than an order of magnitude less massive than its counterpart in the high-purity selection at $z=5$. Despite this difference, the descendant of the least massive galaxy identified in the fiducial selection is nearly 0.6 dex more massive at $z=0$, once again underscoring that relatively low-mass galaxies at $z=5$ can eventually outgrow their more massive coeval counterparts. This difference in subsequent growth appears to be driven by differences in the large-scale environments of the galaxies. The least massive galaxy in the high-purity selection resides within a filament at $z=5$ that later disperses and becomes disconnected from the surrounding matter distribution, limiting its future growth. In contrast, the least massive galaxy in the fiducial selection is embedded in a richer environment populated by numerous nearby halos at $z=5$, providing readily available merger candidates to fuel subsequent growth. 

\begin{figure}
\centering
\hspace*{-0.25in}
\includegraphics[width=3.5in]{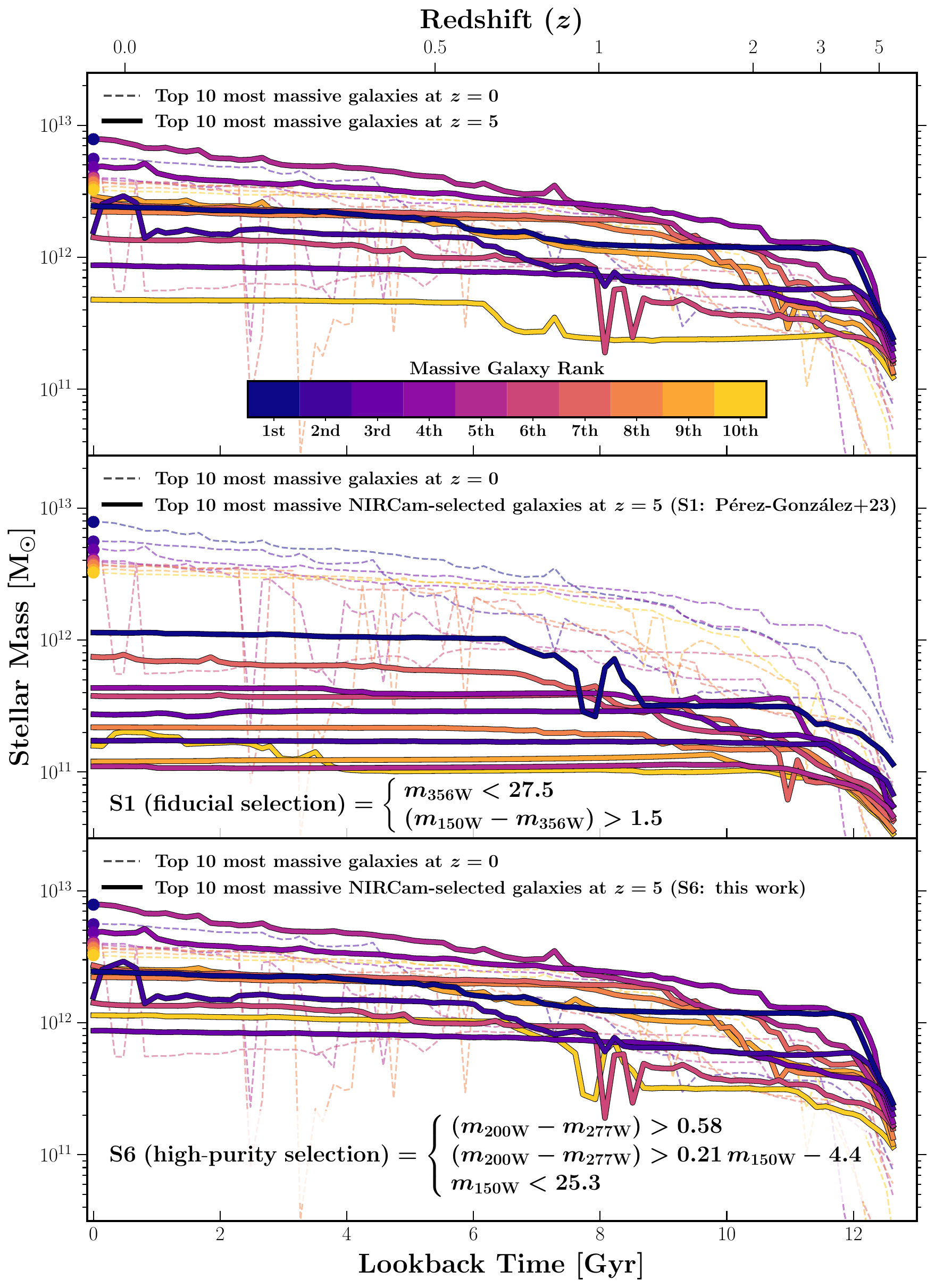}
\caption{Stellar mass growth histories for the ten most massive TNG300 galaxies selected at $z=5$ based purely on stellar mass (top row), using our fiducial NIRCam color selection (middle row), and using our high-purity NIRCam color selection (bottom row). The solid (dashed) lines show the growth (assembly) histories of galaxies that are among the most massive at $z=5$ ($z=0$). The colors of the solid and dashed lines indicate the stellar mass ranking of each galaxy at $z=5$ and $z=0$, respectively. The key takeaways are that (i) the most massive galaxies at early times do not necessarily remain the most massive at late times, and (ii) the high-purity selection more effectively identifies progenitors of massive galaxies than the fiducial selection.}
\label{fig:fig8}
\end{figure}

\begin{figure}
\centering
\hspace*{-0.25in}
\includegraphics[width=3.5in]{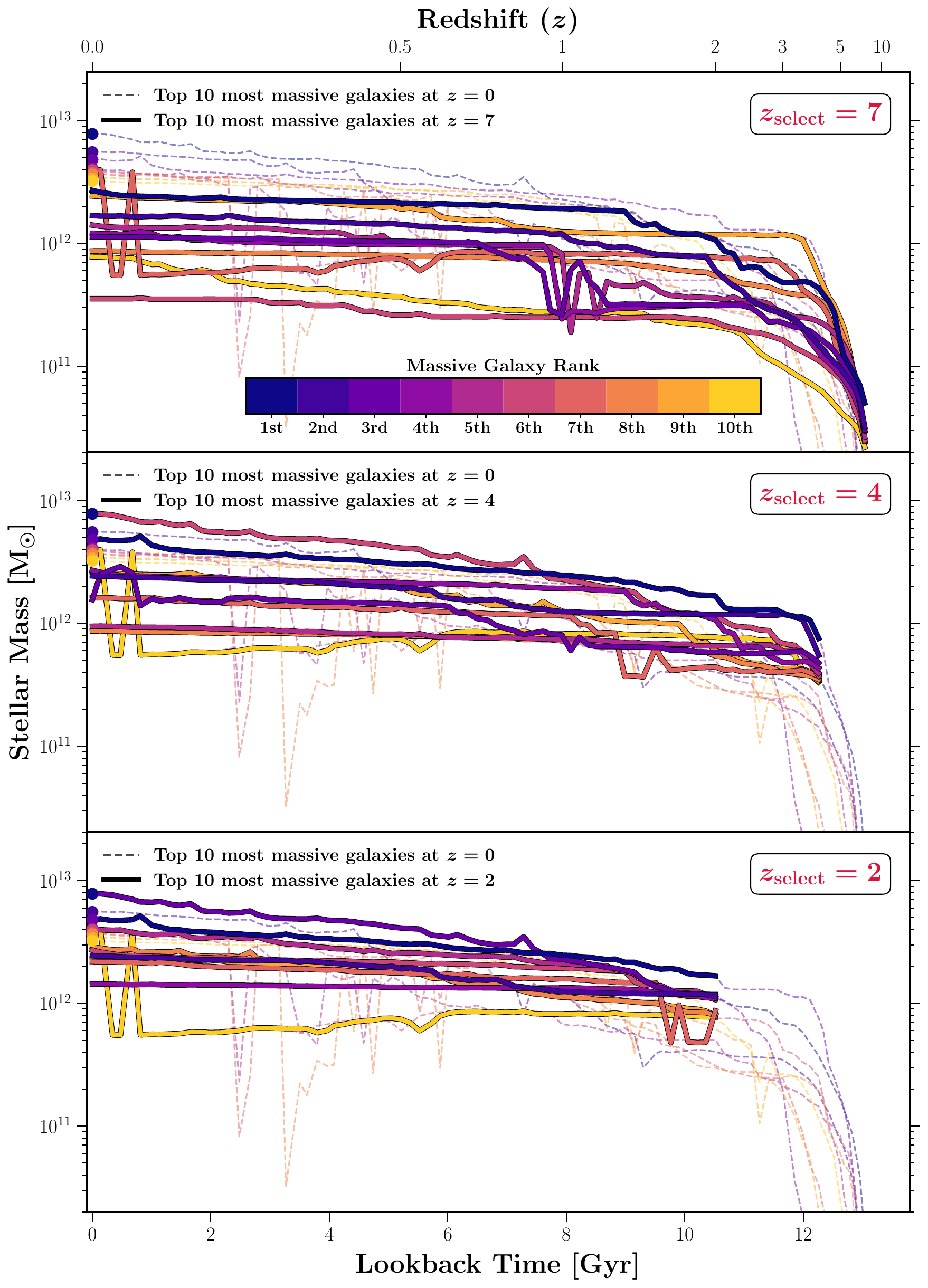}
\caption{Stellar mass growth histories for the ten most massive TNG300 galaxies selected at $z = 7, 4,$ and $2$ (solid lines in the top, middle, and bottom rows, respectively). These growth histories are compared with the assembly histories of the ten most massive galaxies selected at $z=0$, indicated by the dashed lines. The colors of the solid and dashed lines indicate the stellar mass ranking of each galaxy at $z_{\mathrm{select}}$ and $z=0$, respectively. These comparisons reinforce that the most massive galaxies at early times do not necessarily remain the most massive at late times. However, they also show that the overlap increases toward lower redshift, with 4 out of 10 of the most massive galaxies selected at $z=2$ being descendants of the ten most massive galaxies at $z=0$, compared to 1 out of 10 at $z=7$.}
\label{fig:fig9}
\end{figure}

\begin{figure*}
\centering
\hspace*{-0.25in}
\includegraphics[width=7in]{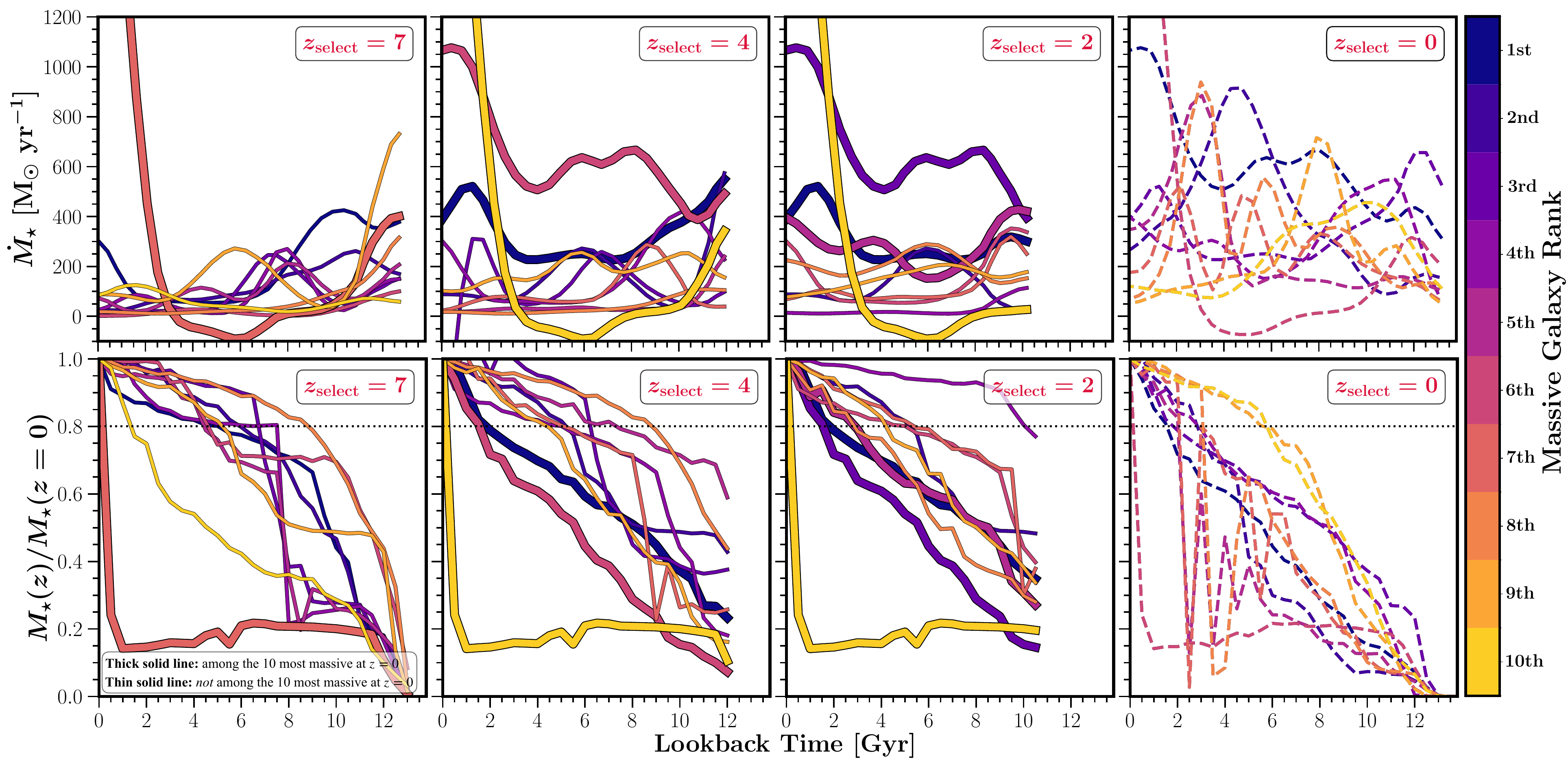}
\caption{\textit{Top row:} Stellar mass growth rates for the ten most massive TNG300 galaxies at $z_{\mathrm{select}} = 7, 4, 2,$ and $0$. \textit{Bottom row:} The fraction of the $z=0$ stellar mass assembled for the ten most massive galaxies at $z_{\mathrm{select}} = 7, 4, 2,$ and $0$. The color bar shows the stellar mass ranking of each galaxy at $z_{\mathrm{select}}$, and the thick (thin) solid lines highlight galaxies that are (\emph{are not}) among the ten most massive at $z=0$. These results show that galaxies selected as most massive at early times typically do not grow into the most massive galaxies at $z=0$; those that do tend to undergo significant late-time stellar mass growth via galaxy mergers.}
\label{fig:fig10}
\end{figure*}

\subsection{Stellar Mass Growth Histories of the Most Massive NIRCam-selected Galaxies}
\label{subsec:4.3}

In this section we perform a comparative analysis of the descendants of high-redshift massive galaxies, focusing on the overall most massive galaxies at $z=5$ in the TNG300 simulation and the most massive NIRCam-selected galaxies identified using both the fiducial and high-purity selections. We restrict our comparison to the ten most massive galaxies in each sample and investigate whether these systems evolve into the most massive galaxies at $z=0$.

The top panel of Fig.~\ref{fig:fig8} compares the stellar mass growth histories of the ten most massive galaxies in the simulation at $z=5$ (solid lines) with the assembly histories of the ten most massive galaxies at $z=0$ (dashed lines), color-coded by their mass rank at $z=5$ and $z=0$, respectively. We find that only two of the descendants of the ten most massive galaxies at $z=5$ are among the ten most massive galaxies at $z=0$, reiterating that high-redshift massive galaxies are not guaranteed to evolve into the most massive galaxies at late times. 

The middle panel compares the ten most massive galaxies identified by the fiducial S1 selection at $z=5$ with the assembly histories of the ten most massive galaxies at $z=0$. We find that none of the galaxies identified by the fiducial NIRCam selection are among the ten most massive galaxies at $z=0$, and only one out of ten reaches a stellar mass above $10^{12}~\msun$ by $z=0$. 

The bottom panel shows the same comparison for the high-purity NIRCam selection. Unlike the fiducial selection, two of the descendants of the ten most massive galaxies in the high-purity sample are among the ten most massive galaxies in the simulation at $z=0$. The high-purity selection is also effective at identifying progenitors of galaxies with stellar masses exceeding $10^{12}~\msun$ at $z=0$, with nine out of ten galaxies satisfying this condition (compared to 8/10 when selecting purely by stellar mass and 1/10 for the fiducial selection).

The results from Fig.~\ref{fig:fig8} highlight that galaxies identified as the most massive at $z=5$ are not guaranteed to evolve into the most massive galaxies at $z=0$ (see \S\ref{subsec:5.2} for a discussion of how this relates to previous studies on the formation of brightest cluster galaxies). In Fig.~\ref{fig:fig9} we extend this analysis by comparing the stellar mass growth histories of the ten most massive galaxies at $z=7$ (top panel), $z=4$ (middle panel), and $z=2$ (bottom panel), against the assembly histories of the most massive galaxies at $z=0$. We find that the overlap between high-redshift massive galaxies and the most massive galaxies at $z=0$ increases as we decrease the selection redshift. Specifically, only 1 out of the 10 most massive galaxies selected at $z=7$ is among the ten most massive galaxies at $z=0$, compared to 3 out of 10 at $z=4$ and 4 out of 10 at $z=2$. Overall, these results demonstrate that massive galaxies identified at high redshift are not generally the progenitors of the most massive galaxies in the universe today, and that such claims should be treated with caution. 

\subsection{Why Most High-Redshift Massive Galaxies Do Not Become the Most Massive Galaxies at z=0}
\label{subsec:4.4}

To investigate why only a subset of high-redshift massive galaxies evolve into the most massive galaxies at $z=0$, we examine the stellar mass growth rates ($\dot{M}_{\star}$), measured in bins of $\sim0.5$ Gyr, for the ten most massive galaxies selected at $z=7,~4,~2$ and $0$ (top row of Fig.~\ref{fig:fig10}). We find that the most massive galaxies at $z\geq2$ that ultimately join the ranks of the ten most massive galaxies at $z=0$ (denoted by the thick solid lines) share a common characteristic: they undergo substantial late-time stellar mass growth during the final $\sim2$ Gyr of the simulation. The rapid stellar mass growth experienced by these galaxies is consistent with them experiencing a major merger (or a series of minor mergers), suggesting that such events are required for high-redshift massive galaxies to grow into the most massive galaxies at $z=0$.

Further evidence that late-time stellar mass growth drives the formation of the most massive galaxies at $z=0$ is provided in the bottom row of Fig.~\ref{fig:fig10}, which shows the cumulative fraction of the final (i.e., $z=0$) stellar mass assembled, also measured in bins of $\sim0.5$ Gyr, for the same galaxies shown in the top row. While many of the most massive galaxies selected at $z\geq2$ \emph{do not} evolve into the most massive galaxies at $z=0$, those that do typically assemble the final $\sim20\%$ of their stellar mass within the last $\sim2$ Gyr of the simulation. This finding further highlights the important role late-time stellar mass growth plays in determining whether or not the most massive high-redshift galaxies evolve into the most massive galaxies at $z=0$.

Taken together with the results from the previous subsections, our findings suggest that the stellar mass and large-scale environment of galaxies selected at $z\geq2$ provide limited predictive power for identifying the progenitors of the most massive galaxies at $z=0$. Instead, whether a massive high-redshift galaxy ultimately joins the ranks of the most massive galaxies at $z=0$ is largely determined by its stellar mass growth at $z<1$, which is inherently difficult to predict for galaxies observed during the first $1-2$ Gyr of cosmic history. Consequently, high-redshift galaxy selections are inherently limited in their ability to identify the progenitors of the most massive galaxies at $z=0$ with high purity. More broadly, our results suggest that interpretations of massive high-redshift galaxies as direct progenitors of the most massive galaxies at $z=0$ should be treated with caution.

\section{Discussion}
\label{sec:Discussion}

\subsection{Diversity in the Descendant Mass of High-Redshift Dark Matter (Sub)halos}
\label{subsec:5.1}
One key takeaway from this analysis is that the most massive high-redshift dark matter halos are not exclusively progenitors of the most massive halos at $z=0$. Instead we find considerable diversity in the $z=0$ descendant halo masses of high-redshift dark matter (sub)halos. Specifically, we find that some of the galaxies occupying the low-mass tail of the dark matter (sub)halo mass distribution at $z=5$ evolve into cluster-mass halos by $z=0$, while some of the most massive halos at $z=5$ fail to reach these halo mass scales. These results are broadly consistent with other theoretical predictions for the descendant halo mass distributions of the most massive high-redshift central galaxies based on constrained simulations~\citep{2023MNRAS.526.2542C} and semi-analytic models~\citep{Nadler23}, which typically span $\sim 10^{12}~\msun$ to $10^{15}~\msun$ in present-day halo mass. However, we find that the range of descendant halo masses for the full range of massive high-redshift galaxies relevant for \textit{JWST} is even wider, with NIRCam color-selected satellites reaching present-day subhalo masses as low as $\sim 10^9~\msun$ in the TNG300 simulation (Fig.~\ref{fig:fig6}, right panel).

These findings have important implications for observational studies of galaxy protocluster candidates, as they suggest that the conventional mass-based methods adopted to estimate the descendant halo mass of these high-redshift galaxy overdensities are potentially unreliable. This has been  explored in recent simulation-based studies of galaxy protoclusters, which generally find that there is not a one-to-one mapping between the most massive structures in the early universe and their present-day descendants \citep[e.g.,][]{Remus23, Lim24, Lim26, Witten26b}. The difficulty in making these connections is further compounded by observational incompleteness, which, as shown in \citet{Baxter25}, significantly impacts the identification of high-redshift galaxy overdensities that are genuine progenitors of present-day clusters. These results highlight the observational challenges of connecting high-redshift overdensities to their present-day descendants and motivate the development of more robust methods for identifying genuine protoclusters, including approaches that leverage simulations to forward model the evolution of observed galaxy overdensities.

\subsection{Brightest Cluster Galaxies Do Not Uniquely Descend from High-Redshift Massive Galaxies}
\label{subsec:5.2}

The most massive galaxies in the present-day universe are the giant elliptical galaxies that reside near the centers of galaxy clusters. These galaxies, known as Brightest Cluster Galaxies (BCGs), have been studied in great detail in part due to their relative proximity; far less is known about their high-redshift progenitors (i.e., proto-BCGs). The prevailing picture of BCG formation is that their stellar populations formed relatively early ($z\gtrsim4$), while the BCGs themselves assembled relatively late ($z<1$) through a series of dry mergers \citep[e.g.,][though see \citealt{Rennehan20} for a scenario in which BCGs in the most massive cluster halos preferentially assemble at early times]{DeLucia07, Jing21, Montenegro23, Vicentin26}. This picture of late BCG assembly is consistent with our findings that the most massive galaxies at $z=0$ are (i) assembled at late times and (ii) not necessarily the descendants of the most massive galaxies at high redshift.

Despite theoretical predictions that BCGs assemble relatively late, a common starting point for observational searches for BCG progenitors involves targeting the brightest (or most massive) galaxy in high-redshift overdensities \citep[e.g.,][]{Ito19, Shi24}. Ignoring the observational challenges involved in identifying genuine galaxy cluster progenitors, the assumption that the most massive galaxy in these structures is the the proto-BCG is not fully supported by this work or previous studies. Using mock HSC-SSP catalogs \citet{Vicentin25a} found that BCG progenitors were correctly identified as the most massive galaxies at high redshift only $\sim65$ per cent of the time. Similar conclusions were reached by \citet{Baxter25}, who found that the most massive galaxies at $z>2$ in protoclusters from the TNG-Cluster simulation \citep{Nelson24} do not necessarily coincide with the progenitors of present-day BCGs.

Nevertheless there is an emerging consensus that high-redshift overdensities, irrespective of whether they ultimately evolve into galaxy clusters, are sites of enhanced stellar mass growth, star formation activity, and merger activity \citep[e.g.,][]{Shimakawa18, Koyama21, Shi21, Forrest24b, Morishita24, Staab24, Sun25, Gururajan25, Giddings2026, Sikorski26}. These conditions are conducive to the formation of massive galaxies at high redshift and have motivated searches for protoclusters using massive and ultra-massive galaxies as biased tracers \citep[e.g.,][]{McConachie21, UrbanoStawinski24, McConachie25}. While comparisons to mock catalogs suggest that such approaches can identify protoclusters with relatively high purity \citep[see][]{Vicentin25b}, our results indicate that the progenitors of present-day BCGs are not necessarily among the most massive galaxies at high redshift. Consequently, proto-BCG searches that focus exclusively on the most massive galaxies may overlook lower-mass candidates with potential to evolve into the most massive galaxy by $z=0$.

\subsection{Impact of Dust Model Assumptions}
\label{subsec:5.3}

The dust-attenuated \textit{JWST}/NIRCam broadband photometry utilized in this analysis was derived by post-processing galaxies from TNG300 using a modified version of the \texttt{SKIRT} full Monte Carlo radiative transfer code \citep{Baes11, Camps13, CampsBaes15, Saftly14}. This procedure inherently involves several assumptions, including those regarding the dust grain size distribution and composition, as well as how the fraction of metals locked into dust evolves with redshift, gas surface density, and galaxy stellar mass.

As explored in \citet{Shen20}, a shortcoming of this model is that it is inconsistent with the observed UV continuum slope ($\beta$) versus $M_{\rm{UV}}$ relation at $z=4-6$ for bright/high-mass galaxies. Introduced in \cite{Calzetti94}, the UV continuum slope $\beta$ encodes the amount of UV light absorbed and scattered by dust, with steeper (shallower) slopes indicating lower (higher) levels of dust attenuation. Therefore, the fact that the massive galaxies in the simulation exhibit steeper UV slopes compared to their observed counterparts at $z>4$ suggests that the amount of dust (or, relatedly, the amount of UV attenuation) in the simulated galaxies is potentially \emph{underestimated}. Given that the NIRCam color criteria presented in \citet{Labbe23}, \citet{Akins23}, and \citet{Gentile24} are designed to identify extremely red (i.e., dusty and/or high-$z$) objects, this shortcoming may be the main reason why we are unable to identify analogs to these galaxies in the simulation.   

In addition to limiting our ability to identify simulated analogs to extremely dusty high-$z$ galaxies, this underlying discrepancy may also impact other conclusions drawn from this analysis. As shown in the top-right panel of Fig.~\ref{fig:fig2} the fiducial NIRCam selection identifies only 1 of the 17 galaxies more massive than $\mstar = 10^{11}~\msun$. Relative to the other 16 massive galaxies, the NIRCam-selected galaxy is $\sim0.5$ mag redder in $m_{\rm{150W}} - m_{\rm{356W}}$. However, since the dust content of massive galaxies is likely underestimated in this simulation, the number of galaxies more massive than $\mstar = 10^{11}~\msun$ selected by the fiducial selection should be interpreted as a lower limit. Therefore, while the number of galaxies identified by the fiducial NIRCam selection with $\mstar > 10^{11}~\msun$ appears low at face value, this estimate should itself be regarded as a lower limit as increasing the amount of UV attenuation for massive galaxies would result in redder $m_{\rm{150W}} - m_{\rm{356W}}$ colors. More broadly, at the high-mass end, the fiducial NIRCam selection likely identifies only the most heavily dust-attenuated galaxies in the simulation.

Similarly, the underestimation of dust attenuation in bright, massive galaxies at high redshift may affect the specific high-purity and high-completeness selections presented in this work for galaxies at $z=5$. Increasing the dust attenuation for the most massive galaxies would effectively shift both these galaxies and the corresponding selection wedges toward the upper right in the color–magnitude space shown in Fig.~\ref{fig:fig4} (i.e., toward fainter $m_{\rm{150W}}$ magnitudes and redder $m_{\rm{200W}} - m_{\rm{277W}}$ colors). Nevertheless, regardless of the magnitude of this shift, the high-purity selection should still outperform the fiducial NIRCam selection as it minimizes contamination from red, low-mass galaxies.

As discussed in \citet{Shen20} the discrepancy between the observed and predicted UV continuum slope versus $M_{\rm{UV}}$ relation at $z=4-6$ likely stems from the simplified treatment of dust in the post-processing procedure, including assumptions regarding the dust grain size distribution and composition adopted from \citep{DraineLi07}. Other assumptions, such as spatially constant dust-to-metal ratios, may also be overly rigid. For example, studies of nearby galaxies have shown that the fraction of metals locked into dust increases with local gas surface density \citep[e.g.,][]{Chiang18, RomanDuval19}. Addressing these simplifying assumptions would require a more sophisticated approach, such as adopting dust population predictions derived self-consistently from simulations that explicitly model dust creation, growth, and destruction in high-redshift galaxies \citep[e.g.,][]{Li19, Choban25, Narayanan26}. However, we reiterate that the adopted NIRCam selections likely identify only the most extreme massive galaxies at high redshift (i.e., the most heavily dust-attenuated systems), while the underlying population satisfying these criteria is almost certainly more numerous.

\section{Summary and Conclusions}
\label{sec:Conclusion}

In this study we leverage synthetic dust-attenuated \textit{JWST}/NIRCam photometry of galaxies in the TNG300 simulation to select analogs of the massive high-redshift galaxy candidates identified in photometric selections from the observational literature. Anchoring our analysis at $z=5$, we assess the effectiveness of five \textit{JWST}/NIRCam photometric selections in recovering similarly massive galaxies in the simulation. We ultimately adopt the NIRCam color selection presented in \citet{PerezGonzalez23} as the fiducial selection because it largely encompasses the galaxy populations recovered by the other selections. We compare the star formation rates, stellar masses, and dark matter halo masses between galaxies indentified by the fiducial NIRCam selection and the progenitors of galaxies more massive than $\mstar = 10^{12}~\msun$ at $z=0$. We also explore the $z=0$ descendants of galaxies in the fiducial NIRCam selection, derive improved NIRCam color-magnitude criteria for efficiently selecting massive galaxies at $z\sim 5$, and examine what factors determine whether a high-redshift massive galaxy evolves into one of the most massive galaxies at $z=0$. 

The main conclusions from our analysis are as follows:
\begin{enumerate}[label=(\roman*)]
\item \textbf{NIRCam selections adopted from the observational literature typically do not recover the most massive galaxies at $\bm{z=5}$ in TNG300:} We find that over half (97 out of 186) of the simulated galaxies identified by the fiducial NIRCam selection at $z=5$ have stellar masses exceeding $10^{10}~\msun$; however, only 1 of the 17 galaxies with stellar masses exceeding $10^{11}~\msun$ is included in the fiducial NIRCam selection (see top row of Fig.~\ref{fig:fig2}). Nevertheless, this should be interpreted as a lower limit since comparisons to observed UV continuum slopes suggest that dust attenuation in the simulated massive galaxies is underestimated.
\item \textbf{NIRCam-selected galaxies in TNG300 are generally not the progenitors of the most massive galaxies at $\bm{z=0}$:} Relative to the progenitors of galaxies more massive than $10^{12}~\msun$ at $z=0$, the NIRCam-selected galaxies at $z=5$ exhibit systematically lower SFRs and reside in less massive dark matter (sub)halos (see bottom row of Fig.~\ref{fig:fig2}). Moreover, of the 186 NIRCam-selected galaxies identified at $z=5$, only 79 survive to $z=0$ without being tidally disrupted. Of these 79, we find that only 8 acquire stellar masses exceeding $10^{12}~\msun$ by $z=0$ (see Fig.~\ref{fig:fig6}).
\item \textbf{Magnitude-dependent NIRCam color selections are most efficient at isolating massive galaxies at $\bm{z\sim5}$ in TNG300:} We explore NIRCam filter combinations that efficiently isolate galaxies more massive than $10^{10.5}~\msun$ at $z\sim5$. We find that optimizing on either purity or completeness favors selection using the $F150W$, $F200W$, and $F277W$ NIRCam filters. These selections, which we label as high-purity (S6) and high-completeness (S7) are described in \S\ref{subsec:3.3}, are effective due to them bracketing the Balmer break at $z=5$ (see Fig.~\ref{fig:fig3}). Additionally, while most current observational selections use strict color cuts, we find that magnitude-dependent color cuts more efficiently identify massive galaxies at $z\sim5$ in the simulation by excluding dusty, low-mass galaxies selected by published observational criteria.
\item \textbf{The most massive simulated galaxies at \boldmath{$z>2$} are not guaranteed to evolve into the most massive galaxies at \boldmath{$z=0$}:} We find that the stellar mass growth of the most massive galaxies at $z>2$ typically plateaus at late times, such that these galaxies do not generally evolve into the most massive galaxies at $z=0$ (see Figs.~\ref{fig:fig8} and ~\ref{fig:fig9}). Instead, membership among the most massive galaxies at $z=0$ is largely determined by merger-driven growth at $z<1$ (see Fig.~\ref{fig:fig10}). Consequently, there appear to be no reliable selections capable of identifying the high-redshift progenitors of the most massive galaxies at $z=0$ with high purity. 
\end{enumerate} 

This work provides insight into the nature and descendants of galaxies identified through NIRCam photometric selections commonly adopted in the literature to isolate high-redshift massive galaxies. Although this analysis likely underestimates the full population of massive NIRCam-selected analogs due to limitations in the adopted dust modeling, future work incorporating more sophisticated and self-consistent dust prescriptions could address this issue. Nevertheless, this study provides a benchmark for the potential range of galaxies identified by contemporary NIRCam photometric selections. This analysis also introduces novel NIRCam photometric selections that could potentially be adopted to efficiently isolate massive galaxies at $z\sim5$, motivating the use of simulations to fully explore and refine color–magnitude selections for identifying high-redshift massive galaxies. Lastly, our results demonstrate that connecting high-redshift massive galaxies to the most massive galaxies at $z=0$ is non-trivial, as this evolutionary link depends strongly on whether or not the descendants of high-redshift massive galaxies undergo substantial late-time merger growth. Upcoming wide-area surveys that map the cosmic web at $z>2$ (e.g., the Nancy Grace Roman Space Telescope) will help establish these evolutionary connections by identifying high-redshift massive galaxies that reside in rich cosmic environments that facilitate merger-driven growth.

\begin{acknowledgments}

We thank Arianna Long and Gourav Khullar for helpful discussions related to this work. We also thank the anonymous reviewer for their thoughtful comments that have enhanced the quality and clarity of this manuscript. DCB is supported by an NSF Astronomy and Astrophysics Postdoctoral Fellowship under award AST-2303800. DCB is also supported by the Future Faculty in the Physical Sciences Postdoctoral Fellowship. ALC is supported by the Ingrid and Joseph W. Hibben endowed chair at UC San Diego. XS would like to acknowledge the support of the NASA theory grant JWST-04814. MV would like to acknowledge the support of the NASA theory grant JWST-04814.

This research made extensive use of {\texttt{Astropy}},
a community-developed core Python package for Astronomy
\citep{Astropy13, Astropy18}.
Additionally, the Python packages {\texttt{NumPy}} \citep{numpy},
{\texttt{iPython}} \citep{iPython}, and
{\texttt{matplotlib}} \citep{matplotlib} were utilized for our data
analysis and presentation. We also make use of the  {\texttt{py-sphviewer2}} \citep{Ben25} python package for visualizing smoothed-particle hydrodynamics data.
In addition, this research has made use of NASA’s Astrophysics Data System Bibliographic Services.

\end{acknowledgments}

\bibliography{citations}{}
\bibliographystyle{aasjournal}

\end{document}